\documentclass{article}

\usepackage{PRIMEarxiv}

\usepackage[utf8]{inputenc} 
\usepackage[T1]{fontenc}    
\usepackage{url}            
\usepackage{booktabs}       
\usepackage{amsmath}        
\usepackage{amsfonts}       
\usepackage{nicefrac}       
\usepackage{microtype}      
\usepackage{lipsum}
\usepackage{fancyhdr}       
\usepackage{graphicx}       
\graphicspath{{media/}}     
\usepackage{endnotes}
\usepackage{tabularx}

\usepackage{comment}
\usepackage{subfigure, epsfig}
\usepackage[numbers]{natbib}
\usepackage[hidelinks]{hyperref}  

\title{Inferring Inventory Dynamics from Supply Chain Networks: A Graph Learning Approach with Autonomous Validation}

\author{
  Tiancheng Gao$^{1}$,\ \ Xiang Zhang$^{2}$,\ \ Wei Lan$^{3}$,\ \ Bin Liu$^{4}$ \\
  \\
  $^{1,3,4}$Southwestern University of Finance and Economics, School of Statistics and Data Science, \\
  $^{2}$Southwestern University of Finance and Economics, School of Finance, \\
}

\begin{document}
\maketitle

\begin{abstract}
Supply-demand mismatch represents a fundamental challenge in supply chain management, yet its direct measurement remains particularly elusive for small and medium-sized enterprises (SMEs).These firms typically lack systematic inventory records, leaving labeled training data critically scarce. Conventional supervised learning methods rely heavily on labeled samples, rendering them ill-equipped to reliably validate firm-level predictions—in terms of both predictive reliability and operational utility—under such data-scarce conditions.
To resolve this unlabeled-data dilemma, we develop a multi-agent semi-supervised inference framework that reframes the label-scarcity problem as a structured, collaborative task distributed across specialized agents. We first construct a production-function-constrained graph machine learning model that infers firm-level inventory changes directly from supply chain network topology. A dedicated econometric validation agent then concurrently loads five econometric models—spanning spatial spillovers, dynamic persistence, causal direction, shock transmission, and supply-demand forecasting—to generates structured economic evidence from complementary dimensions. An expert review agent synthesizes the structured econometric evidence and produces a unified consistency assessment by resolving cross-agent inconsistencies.
Empirical results demonstrate stable predictive performance on inventory-change forecasting tasks. Multi-agent econometric validation further confirms that predicted inventory dynamics align closely with established economic theory in terms of causal structure and network transmission mechanisms. Critically, the proposed agent framework enables effective prediction verification even when ground-truth observations are unavailable.
\end{abstract}

\keywords{Supply-Demand Mismatch \and Multi-Agent Reasoning \and Inventory Forecasting \and Production Function \and Graph Neural Networks}
\section{Introduction}\label{sec:Intro}
In today's highly complex and interdependent industrial and supply chains, the dynamic alignment of supply and demand is essential to the efficient functioning of the economic system.
Recent global disruptions have demonstrated that supply-demand mismatch has emerged as a major threat to supply-chain stability.
The COVID-19 pandemic offers a compelling illustration.
According to the World Trade Organization, global merchandise trade volume fell by 5.3\% in 2020, while commercial services trade declined by 21\% year over year, reflecting a simultaneous contraction in global production, transportation, and demand\endnote{World Trade Organization, ``World Trade Primed for Strong but Uneven Recovery after COVID-19 Pandemic Shock,'' March 31, 2021, \url{https://www.wto.org/english/news_e/pres21_e/pr876_e.htm}, accessed May 4, 2026.}.
At the same time, the New York Fed's Global Supply Chain Pressure Index reached a historical peak at the end of 2021, indicating a sharp short-term deterioration in transportation costs, delivery times, and manufacturing supply constraints\endnote{Federal Reserve Bank of New York, ``New York Fed Launches Global Supply Chain Pressure Index,'' May 18, 2022, \url{https://www.newyorkfed.org/newsevents/news/research/2022/20220518}, accessed May 4, 2026.}.
Subsequently, the fluctuations in energy and commodity prices triggered by the Russia-Ukraine conflict further exposed the energy, chemical, transport, and manufacturing sectors to new cost pressures and supply uncertainties\endnote{World Bank, ``Food and Energy Price Shocks from Ukraine War,'' April 26, 2022, \url{https://www.worldbank.org/en/news/press-release/2022/04/26/food-and-energy-price-shocks-from-ukraine-war}, accessed May 4, 2026.}.
These developments demonstrate that supply-demand mismatch in modern industry chain is not merely an isolated business management issue; rather, it is deeply intertwined with global value-chain specialization, dependence on critical inputs, and the transmission of shocks through supply networks.

In practice, a firm's production, procurement, and sales processes do not occur simultaneously.
Demand can shift rapidly in response to market changes, external shocks, or expectation revisions, while supply adjustments are constrained by production schedules, procurement cycles, capacity limitations, and information lags.
Consequently, supply capacity frequently fails to keep pace with demand fluctuations, manifesting at the firm level as inventory imbalances.
This misalignment is particularly pronounced in modern supply-chain networks.
Disruptions such as supply shortages, demand contractions, and delayed production adjustments can propagate through the network and generate systemic fluctuations across the entire chain.
Supply-demand mismatch is therefore not merely an internal managerial problem, but a systemic phenomenon rooted in network interdependence.

Supply-demand mismatch is inherently difficult to observe directly, as both supply and demand are dynamic.
A firm's effective supply capacity depends not only on procurement inputs but also on its own production efficiency, and realized transactions capture only the outcome of the supply-demand matching process without revealing how mismatches arise, accumulate, or propagate through the supply-chain network.
Relying solely on transaction data therefore makes it difficult to identify the true sources of supply-demand imbalances.
By contrast, inventory changes provide a more direct signal: when effective supply exceeds demand, unsold output accumulates as inventory; when demand exceeds effective supply, firms draw down inventory to fulfill orders.
Inventory thus serves as a key state variable linking production, sales, and supply-demand mismatch, reflecting the cumulative imbalance over time.
However, inventory data are largely unavailable for small and medium-sized enterprises (SMEs), making it particularly challenging to study supply-demand mismatch based on inventory changes.

Existing research has approached supply-demand relationships primarily through three streams: demand forecasting, inventory management, and supply chain network analysis.
Demand forecasting studies use historical order or transaction data to predict future demand, thereby assisting firms in making production and procurement decisions.
Early work relied largely on time-series models to characterize demand dynamics \citep{winters1960, roberts1982, gardner1985}, whereas more recent studies increasingly employ machine learning methods to capture nonlinear patterns and complex fluctuations in demand \citep{harahap2025, celestin2025, dasilva2026}.
Although this research enhances the ability to identify demand-side information, the core task remains understanding the supply-demand relationship from the demand side alone.
Because sales and order data typically represent the outcome of an already completed matching process, demand forecasting alone cannot adequately reveal structural imbalances arising from different sources.

In contrast, research on inventory management examines how firms determine ordering strategies, stock levels, and replenishment frequencies, with the goal of reducing inventory costs and improving service levels.
Such studies often assume that demand is known or predictable and include classical models such as the economic order quantity model, reorder-point policies, and safety-stock frameworks.
Subsequent research has further integrated forecasting models and optimization methods to enhance the dynamic adaptability of inventory decisions \citep{olaniyi2026, sagaert2025, zizka2026}.
This literature demonstrates that inventory is a central instrument for coping with demand uncertainty; however, it typically treats inventory as a management target or a decision outcome rather than as a state representation of dynamic deviations between supply and demand.
As a result, such studies are more concerned with how to optimize inventory levels than with how inventory changes reflect underlying supply-demand mismatch.

Research on supply chain networks suggests that upstream and downstream relationships shape how supply-demand fluctuations propagate across firms.
Studies on production networks and input-output networks indicate that local shocks can be amplified or attenuated through interconnections among firms or industries, thereby influencing aggregate economic fluctuations \citep{acemoglu2012, carvalho2021}.
More recently, graph neural networks (GNNs) have been applied to learn structural dependencies in supply chain networks and have been employed for tasks such as relationship prediction, supplier recommendation, industry classification, and dynamic network analysis \citep{wu2023, huang2023, tu2024}.
Although these methods are effective at capturing complex inter-firm interconnections, most existing studies focus on network topology, node-link optimization, or related structural tasks, and rarely incorporate firms' production processes, supply capacity, and inventory evolution into a unified analytical framework.
Consequently, such models may capture network dependence but cannot explain the origins of supply-demand mismatch.

As the foregoing review illustrates, the existing literature provides an important foundation for understanding supply-demand relationships, yet exhibits several critical limitations.
First, demand forecasting studies focus predominantly on the demand side and therefore cannot identify the sources of disequilibrium.
Second, inventory management research treats inventory primarily as an optimization target rather than as a state variable reflecting supply-demand mismatch.
Third, supply-chain network research can map inter-firm relationships but lacks an explicit economic mechanism to explain how mismatches form and propagate.
More importantly, most existing studies assume that inventory data are directly observable.
To the best of our knowledge, no existing approach addresses the widespread unavailability of inventory data for SMEs.

To address these research gaps, our study proceeds as follows.
First, given the inherent difficulty of directly observing supply-demand mismatch, we treat inventory changes as the cumulative outcome of dynamic discrepancies between effective supply capacity and market demand.
This approach converts an otherwise unobservable quantity into measurable inventory dynamics, circumventing the observability problem and laying the groundwork for subsequent empirical analysis.
Second, we provide a rigorous economic interpretation of the mechanism underlying mismatch formation.
We employ a production function to characterize how upstream inputs are transformed into effective supply capacity, while an inventory update equation establishes that inventory changes are equivalent to the gap between effective supply capacity and market demand.
Third, considering the limited availability of inventory data for SMEs, direct validation of the proposed prediction method faces inherent constraints.
To address this, we evaluate the predicted inventory changes through an economics-based consistency framework, verifying whether the predictions can effectively explain the formation, accumulation, and transmission of supply-demand mismatch across supply-chain networks.

The contributions of this study can be summarized along three dimensions.
First, we conceptualize inventory changes as a state representation of supply-demand mismatch.
By interpreting inventory changes as the outcome of dynamic divergence between supply capacity and market demand, this study provides a more meaningful characterization of supply-demand mismatch in industrial and supply chain networks.
Second, we incorporate a production function and an inventory update equation into a graph neural network framework to improve economic interpretability.
By integrating industrial and supply chain network representations with the estimation of firms' effective supply capacity and the inventory update equation, the proposed model captures network dependence while remaining consistent with economic logic.
Third, in the absence of inventory labels, we enhance the credibility of prediction results through a multi-agent economic consistency validation framework.
It evaluates whether inventory forecasts are consistent with inventory cycles and supply chain transmission patterns from multiple economic perspectives, thereby providing a feasible approach for validating predictions in unlabeled settings.

Overall, this study begins from the practical problem of supply-demand mismatch and integrates inventory dynamics, industrial and supply chain network structures, and economic consistency testing into a multi-agent semi-supervised inference framework.
The objective is not only to predict firm-level inventory changes accurately but also to ensure that those predictions can be justified by economic reasoning.
Through this approach, the study offers a new perspective for understanding supply-demand mismatch and its network transmission mechanisms in industrial supply chains.

\section{Related Work}

\subsection{Supply-Demand Mismatch}

Supply-demand mismatch is a central source of instability in supply chains and industry chains.Firms face concurrent challenges, including demand fluctuations, supply constraints, production adjustment delays, and information asymmetry. Mismatch arises when supply capacity fails to align with market demand. In supply networks, these imbalances occur not only within individual firms but also across trading relationships, where local fluctuations can propagate into broader network instability.

Existing studies explain the formation of supply-demand mismatch from the perspectives of inventory cycles, information transmission, and supply chain shocks. \citet{kitchin1923} points out that firms respond to market information with delays, which slow both information transmission and production adjustment and thereby generate short-term economic fluctuations. \citet{metzler1941} further explains business fluctuations from the perspective of the inventory cycle, arguing that firms adjust production and target inventory levels in response to changes in demand. \citet{blinder1991} emphasize that inventory investment is an important variable for understanding output fluctuations. Their findings suggest that inventory changes are not merely operational outcomes but also reflect the dynamic adjustment between supply and demand. Similarly, \citet{bils2000} show that the dynamic relationship between inventories and sales captures adjustment processes over the business cycle. Together, these studies suggest that inventory is not only an operational outcome but also a state variable that reflects the dynamic deviation between supply capacity and market demand.

In the field of supply chain management, supply-demand mismatch is more commonly discussed in relation to information distortion, demand amplification, and upstream-downstream relationships. \citet{lee1997} establish that demand signals amplify as they move upstream, causing the bullwhip effect. Revisiting this from a network perspective, \citet{osadchiy2021} show that changes in a supplier's customer base affect aggregate demand volatility. \citet{qu2021} argue that vertical contracts and wholesale pricing influence production and sales volatility. Furthermore, \citet{chu2017} and \citet{candogan2024} examine how demand forecasting, information sharing, and information design shape upstream capacity decisions. These studies indicate that supply-demand mismatch is driven not only by demand uncertainty but also by information transmission, capacity decisions, pricing mechanisms, and the structure of supply chain relationships.

Despite its importance, supply-demand mismatch is difficult to observe directly. Sales and transaction flows are realized outcomes that cannot clearly distinguish among insufficient demand, supply constraints, or transaction disruptions. For instance, declining sales could stem from either shrinking demand or upstream shortages. Consequently, relying solely on realized flow variables fails to capture the true state of mismatch and its dynamic accumulation. By contrast, inventory changes accumulate the deviations between supply capacity and market demand. Therefore, using inventory changes as a state representation of supply-demand mismatch provides a dynamic adjustment perspective on firm imbalances, overcoming the limitations of single-period transaction outcomes.

\subsection{Inventory Forecasting}
Inventory forecasting is an important problem in supply chain management and operation. Traditional studies on inventory management typically assume that demand is known or predictable and therefore focus on ordering policies, replenishment rules, safety stock, and inventory cost control. With the development of data-driven methods, machine learning, deep learning, and online learning approaches have been increasingly introduced into inventory management. \citet{qi2022} propose an end-to-end model that learns replenishment decisions directly from historical data, bridging the gap between forecasting and optimization in the traditional predict-then-optimize framework. \citet{li2023} show that under inventory constraints and external information, firms must dynamically balance demand learning with inventory depletion. \citet{chen2024} develop an online learning algorithm for lost-sales systems facing unknown demand and supply distributions. \citet{zhang2025} optimize multi-stage serial inventory systems using sample average approximation.

Beyond operational efficiency, recent studies have broadened the economic interpretation of inventory. \citet{wu_lai2022} find that managerial incentives can induce abnormal inventory holdings, indicating drivers beyond pure operations. \citet{hsu_wu2024} demonstrate that inventory also functions as a financial instrument to exploit arbitrage opportunities and alleviate financing constraints. \citet{iancu2017} show that inventory liquidation and replenishment flexibility influence financing costs and risk-taking incentives in debt contracts. Collectively, these studies highlight that inventory serves not only as an operational variable but also reflects financing constraints, governance incentives, and market shifts.

Nevertheless, the existing literature on inventory forecasting leaves two important limitations. First, most studies treat inventory as an outcome variable following demand forecasting or replenishment optimization, with the primary objective of improving inventory management efficiency. Relatively few studies interpret inventory dynamics from the perspective of supply-demand mismatch. Second, existing methods usually assume that inventory data are directly observable and can be used as supervised labels for model training and evaluation. In real industry chain, especially among SEMs, inventory data are often missing or incomplete. Under such circumstances, it is necessary to explore a new strategy for inferring and validating inventory changes.

\subsection{Supply Chain Networks}
Studies of supply chain networks examine trading relationships, input-output structures, and shock propagation, demonstrating how localized shocks spread through linkages to induce aggregate fluctuations. \citet{acemoglu2012} trace these aggregate fluctuations to microeconomic shocks at central network nodes. Using natural disaster evidence, \citet{barrot2016} and \citet{carvalho2021} illustrate how supplier disruptions propagate downstream. Additionally, \citet{antras2012} quantify industry upstreamness in global production. Together, these studies suggest that the supply-demand state of a firm is determined not only by internal operations but also by the position of the firm in industrial and supply chain networks.

Graph neural networks are increasingly applied to model these complex dependencies by integrating firm-level features with network structure. For instance, \citet{wu2023} leverage trading relationships for industry classification, while \citet{tu2024} apply graph representation learning to supplier recommendation to capture latent inter-firm cooperation. Furthermore, \citet{huang2023} provide a temporal graph benchmark for dynamic network modeling. These studies demonstrate the efficacy of graph neural networks in capturing dynamic relationships and upstream--downstream information transmission in supply chains.

However, existing graph neural network methods face notable limitations in predicting inventory dynamics. Most models rely on structural and node features but lack explicit economic constraints. Specifically, by using transaction flows directly as inputs, they fail to distinguish among upstream input scale, transformation efficiency, and effective supply capacity. Consequently, despite achieving high predictive accuracy, it remains difficult to ascertain whether these models align with inventory formation mechanisms and supply-demand mismatch logic. Furthermore, the reliance on supervised learning limits the direct evaluation of these methods for small and medium-sized enterprises, where inventory labels are often missing or incomplete.

\section{Research Hypotheses and Variable Definition}

\subsection{Research Hypotheses}

We study supply-demand mismatch from integrated perspectives of industry chain and supply chain. The production and sales activities of firms do not occur in isolation. Instead, they are embedded in upstream-downstream specialization and transaction networks. The industry chain reflects input-output relationships across industries. These relationships are relatively stable and indicate the position of a firm within the broader production system. By contrast, the supply chain captures dynamic transaction relationships among firms and records the sources of input procurement and the destinations of product sales. Together, these two dimensions constitute the external operating environment of the firm.

Supply-demand mismatch, however, is determined not only by procurement inflows and sales outflows but also by the ability of a firm to convert inputs into marketable output. We define effective supply capability as the marketable supply generated when upstream inputs are transformed through the production process. For example, two firms with similar procurement volumes may still generate very different levels of effective supply if their production efficiency differs.

Under this framework, a firm's supply-demand condition is jointly determined by upstream inputs, production transformation, and downstream demand. Specifically, upstream suppliers provide intermediate goods, while downstream customers determine market absorption. Because upstream inputs do not directly equate to actual supply, firms must convert them into effective marketable output. When this effective supply fails to match downstream demand, the resulting gap drives inventory dynamics.

Based on this conceptual setting, we propose the following core hypotheses. 
First, transaction relationships map a firm's supply-demand environment. Supply-demand relationships are interconnected through the transaction network. Upstream transaction flows reflect the input base, while downstream flows capture market demand. 
Second, effective supply capability is generated by transforming upstream inputs through a production function, and thus does not directly equate to observed procurement flows. 
Third, inventory changes are determined by the gap between effective supply capability and downstream demand, reflecting the dynamic divergence between supply and demand. 

\subsection{Variable Definition}
In our framework, industry classification, listing status, and seasonal characteristics describe a firm's basic attributes and temporal context; therefore, we treat them as exogenous input variables. Industry classification reflects a firm's position in the industry chain. Listing status proxies for firm size, operational stability, governance quality, and financing capacity, while seasonal features control for semiannual temporal variations.

Interfirm transactions, upstream procurement, and downstream sales serve as directly observable behavioral variables. Specifically, upstream procurement characterizes the input base, whereas downstream sales capture realized product demand. Because underlying supply-demand conditions drive these transactions, we do not interpret them as strictly exogenous shocks. Instead, they serve as observable manifestations of a firm's external supply-demand environment and as critical inputs for inferring inventory dynamics.

Production efficiency, effective supply capability, and changes in inventory serve as endogenous variables to be inferred by the model. Specifically, production efficiency captures the capacity of the firm to convert upstream inputs into outputs. Building upon this, effective supply capability denotes the volume of marketable output generated for a given level of upstream inputs and production efficiency. Subsequent changes in inventory are then determined by the discrepancy between this effective supply capability and downstream demand. Consequently, inventory is not an exogenous outcome, but an endogenous state variable jointly shaped by upstream inputs, production transformation, and downstream demand.

Drawing upon these definitions, the relationships among the variables can be summarized as follows: exogenous attribute variables characterize the basic position and temporal environment of the firm; observable transaction variables depict the external supply and demand conditions of the firm; and endogenous latent variables capture the production transformation process and the resulting inventory dynamics. Collectively, these components establish the foundation for modeling supply-demand mismatch within industrial and supply chain networks.

\subsection{Research Approach}

In summary, the analytical process of this research proceeds sequentially.
(1) we characterize the industrial-chain and supply-chain ecosystem in which firms operate using disaggregated industry information and observed interfirm transaction relationships; 
(2) we use a production function to transform upstream input factors into effective supply capability; 
(3) we represent inventory change as the difference between effective supply capability and downstream market demand; 
(4) under conditions of incomplete inventory observability, we use a graph neural network to predict firm-level inventory changes and then econometric multi-agent to assess whether the predictions are consistent with basic economic logic.

Through this comprehensive framework, we integrate the positions of firms within the industry chain, transactions across the supply chain, the production function, inventory dynamics, and supply-demand mismatches into a unified analytical mechanism. As prior literature suggests, traditional approaches to inventory analysis often struggle to capture the complex interplay of relational dependencies, such as supplier-customer linkages, alongside spatiotemporal dynamics, including network-based demand propagation and seasonal fluctuations. Such limitations inherently increase the risk of inventory overhangs or shortages. By contrast, our framework builds on graph machine learning and multi-agent reasoning, providing a feasible approach for identifying supply-demand mismatches in settings with missing inventory information.

\section{Production-Constrained Graph Neural Networks for Inventory Changes Prediction}

\subsection{Firm Production Process}
A firm's production process dynamically transforms input factors into marketable products. The production function characterizes this capability, serving as the fundamental link between inputs---such as labor, capital, and intermediate goods---and outputs. In macroeconomic and industrial research, these functions are routinely employed to assess productivity, capacity utilization, and supply capability. In a supply chain context, procurement inflows act as intermediate inputs, while effective output depends on transformation efficiency. We define the effective production capacity of firm $i$ as:
\begin{equation}
    Q_i(t) = S_i(t) \cdot \left(1 + \alpha_i(t)\right),
    \label{eq:effective_capacity}
\end{equation}
where \(S_i(t)\) denotes the total procurement inflow of firm \(i\) in period \(t\); \(\alpha_i(t)\) denotes the effective production rate of firm \(i\) in period \(t\), reflecting the efficiency with which the firm transforms inputs into outputs. \(Q_i(t)\) can be interpreted as the firm’s effective supply capacity, that is, the level of output that the firm can provide under given input conditions. Therefore, inventory changes are essentially driven by the difference between effective supply capacity and demand.

It should be noted that the parameter \(\alpha_i(t)\) captures efficiency deviations stemming from unmodeled operational factors, including labor, capital utilization, managerial capability, and production losses during the production process. When $-1 < \alpha_i(t) < 0$, production efficiency falls below the benchmark, indicating suboptimal operations. Furthermore, when $\alpha_i(t) < -1$, the marginal contribution of production becomes negative; due to severe inefficiency or resource misallocation, additional inputs reduce net inventory. This specification reflects the real-world phenomenon where increased investment exacerbates losses.

In addition, changes in firm inventory are also affected by market demand. On the one hand, production capacity determines the effective supply that the firm can provide; on the other hand, market demand determines the extent to which its products are actually absorbed. When supply exceeds demand, surplus output is transformed into inventory accumulation. Conversely, when demand exceeds supply, inventory is consumed to make up for the supply shortage. Therefore, inventory reflect not only an operating outcome but also the dynamic adjustment process between supply and demand.

However, observed supply and demand alone cannot fully explain inventory dynamics. Classic research shows that information delays, shrinkage, and measurement errors create discrepancies between recorded and actual inventory \citep{iglehart1972}. \citet{dehoratius2008} document pervasive inventory record inaccuracies, confirming that accounting-based flows fail to capture the true inventory state. Furthermore, \citet{chehrazi2025} demonstrate that transaction errors and unobserved shrinkage distort dynamic inventory assessments. Accordingly, we introduce a dynamic adjustment term, $\epsilon_i(t)$, to capture unobserved deviations affecting inventory.

From the perspective of the inventory cycle, we define supply-demand mismatch as the dynamic divergence between supply capability and realized demand, and we interpret inventory changes as the cumulative outcome of this divergence. Under this framework, inventory changes of firms can be formalized as a state evolution process that satisfies flow conservation. We model this process as a generalized inventory state transition equation \citep{chiang2023, chen2024}. Specifically, $I_i(t)$, that is the inventory of the firm $i$ at time $t$ is jointly determined by the inventory at the beginning period, effective supply capability, and the outflows generated by sales,
\begin{equation}
    I_i(t) = I_i(t-1) + Q_i(t) - D_i(t) + \epsilon_i(t),
    \label{eq:inventory_state}
\end{equation}
where \(I_i(t-1)\) is the ending inventory of firm \(i\) in period \(t-1\); \(D_i(t)\) denotes the total demand or sales outflow of firm \(i\) in period \(t\); and \(\epsilon_i(t)\) represents a dynamic adjustment term used to explain latent operational behaviors that cannot be directly observed.

Taking the first-order difference of the above equation yields the inventory changes:
\begin{equation}
    \Delta I_i(t) = I_i(t) - I_i(t-1)
    = Q_i(t) - D_i(t) + \epsilon_i(t),
    \label{eq:inventory_change}
\end{equation}
this form directly reveals the economic meaning of inventory: when inventory changes that cannot be directly observed are not considered, the inventory changes \(\Delta I_i(t)\) is equal to the difference between effective supply and demand. Therefore, inventory dynamics reflect the cumulative result of supply-demand mismatch over time. An increase in inventory indicates excess effective supply, while a decrease in inventory indicates that market demand exceeds supply capacity.

\subsection{Definition of the Supply Network}

The inventory changes of a firm remain fundamentally driven by the discrepancy between effective supply and market demand. However, these conditions do not operate in isolation. In supply chains, upstream output serves as downstream input, while downstream demand reciprocally shapes upstream production. Consequently, a firm-level supply-demand mismatch is not merely an internal operational outcome, but a manifestation of network-wide interactions. Therefore, firms must be modeled as embedded entities within the broader supply network.

In feature design, we integrate two types of information. On the one hand, dynamic transaction flows and network structure in the supply chain are used to characterize short-term fluctuations in supply and demand. On the other hand, industry affiliation and position of the industry in the industry chain are further introduced as static structural features of firms, providing long-term structural constraints. Accordingly, under the structural constraints of the industry chain, we characterize the evolution of supply-demand mismatch through dynamic interactions in the supply chain.

To model the network structure of supply and demand relationships between firms, 
we formalize the supply chain as a dynamic directed graph $G(t) = (\mathcal{V}, \mathcal{E}(t), \mathbf{X}(t))$. 
Here, $t \in \{0,1, 2, \dots, 16\}$ and $\mathcal{V} = \mathcal{V}_{listed} \bigcup \mathcal{V}_{SMEs}$ denotes the set of $n$ firm nodes.
The inventory information of listed firms set $\mathcal{V}_{listed}$ is observed and can provide supervision signals for model estimation, whereas the inventory information of SMEs set $\mathcal{V}_{SMEs}$ is unobserved and must be inferred from supply chain interactions. 
$\mathcal{E}_t$ denotes the set of directed edges at time $t$.
Each edge $(j \rightarrow i)$ indicates that firm $i$ purchases raw materials and intermediate goods from firm $j$. $\mathbf{X}(t) \in \mathbb{R}^{n \times d}$ is the node feature matrixbutes of firms at time $t$. 

Building on this, we introduce the primary notation used in this section. 
For any firm node $i$, its sets of upstream suppliers and downstream customers are respectively defined as:
\begin{equation}
\mathcal{N}_{\mathrm{in,i}}(t)
=
\{j \in \mathcal{V} \mid j \to i \in \mathcal{E}(t)\},
\end{equation}
\begin{equation}
\mathcal{N}_{\mathrm{out,i}}(t)
=
\{j \in \mathcal{V} \mid i \to j \in \mathcal{E}(t)\}.
\end{equation}

Let $T_{ji}(t)$ denote the value of a transaction from firm $j$ to firm $i$ at time $t$.
Based on equation~\eqref{eq:inventory_state}, we compute inflows and outflows of firm $i$ at time $t$ by aggregating transactions over its upstream and downstream neighborhoods:
\begin{equation}
    S_i(t) = \sum_{j \in \mathcal{N}_{in,i}(t)} T_{ji}(t), \quad D_i(t) = \sum_{j \in \mathcal{N}_{out,i}(t)} T_{ij}(t).
\end{equation}

In this way, upstream transactions $S_i(t)$ determine the input base available for production, whereas downstream transactions $D_i(t)$ measure the extent to which the output of firm $i$ is absorbed by the market.

For node features, let the feature vector of firm $i$ at time $t$ be denoted by $\mathbf{x}_i(t)$, which corresponds to the $i$-th row of $\mathbf{X}(t)$, that is, $x_i(t) = \mathbf{X}_i(t)$.
The feature vector $x_i(t)$ comprises the following five components: (1) firm identity; (2) industry category; (3) listing status; (4) a seasonal indicator distinguishing the first and second halves of the fiscal year; and (5) a momentum term capturing the lagged inventory change $\Delta I_i(t-1)$.
Together, these components enable the feature representation to capture both cross-sectional heterogeneity across firms and temporal dynamics in inventory evolution.We transform the raw features into the feature fusion layer to obtain the initial node representation:
\begin{equation}
{H}^{(0)}_{i}(t)=\sigma ({x}_i(t);\mathbf{W}_f,\mathbf{b}_f),
\end{equation}
where $\mathbf{h}^{(0)}_{i}(t) \in \mathbb{R}^d$ denotes the initial hidden representation of firm $i$ at period $t$; $d$ represents the hidden space dimension after feature mapping; $\sigma(\cdot)$ is the ReLU function.

Using equation~\eqref{eq:effective_capacity} - \eqref{eq:inventory_change}, we can obtain the inventory level of firm $i$ at time $t$ ($I_i(t)$), with its changes  $\Delta I_i(t)$. 
Based on our understanding of the mismatch between supply and demand, the inventory changes $\Delta I_i(t)$ can be interpreted as a direct measure of supply-demand mismatch and will be modeled as the core variable in the subsequent analysis.

\subsection{Directed Graph Neural Networks}\label{sec:DGNN}

To capture the directionality and asymmetry of information transmission in supply chains, we draw on the idea of asymmetric representations in directed graph neural networks \citep{tan2022}. Compared with undirected edges, directed edges carry richer relational information; therefore, information flows in different directions should be treated differently during propagation. Accordingly, each firm node is represented as playing two behavioral roles simultaneously: a ``receiver'' that obtains inputs from upstream firms and a ``sender'' that transmits output to downstream firms.

Based on this perspective, for each node \(i\) at time \(t\), we introduce two types of embeddings: a receiving embedding and a sending embedding, denoted by $\mathbf{r}_i^{(l)}(t)$ and $\mathbf{o}_i^{(l)}(t)$, respectively. These embeddings characterize the capability of a firm to acquire inputs and transmit output within the supply chain. The core of this modeling strategy is to decompose information flow in the supply chain into two directional processes with clear economic meaning, rather than simply aggregating neighborhood features.

In implementation, we adopt a stepwise asymmetric message-passing mechanism. 
In each propagation layer, a node first aggregates information from its upstream neighbors to update the receiving embedding and then aggregates information from its downstream neighbors to update the sending embedding. 
This sequence corresponds closely to the actual process in supply chains:
firms first obtain production inputs through upstream relationships and then
distribute output through downstream relationships, thereby forming a two-way
aggregation of supply and demand. Formally, this process can be written as:
\begin{align}
{M}_{i,in}^{(l)}(t)
&=
\text{AGG}\left({R}_j^{(l-1)}(t), j \in \mathcal{N}_{in}(i)\right), \\
{M}_{i,out}^{(l)}(t)
&=
\text{AGG}\left({O}_j^{(l-1)}(t), j \in \mathcal{N}_{out}(i)\right), \\
{R}_i^{(l)}(t)
&=
\sigma\left(
\mathbf{W}_r^{(l)} ; \quad
{R}_i^{(l-1)}(t),\quad
{M}_{i,in}^{(l)}(t)
\right) ,\\
{O}_i^{(l)}(t)
&=
\sigma\left(
\mathbf{W}_o^{(l)} ; \quad
{O}_i^{(l-1)}(t),\quad
{M}_{i,out}^{(l)}(t)
\right),
\end{align}
where ${M}_{i,in}^{(l)}(t)$ and ${M}_{i,out}^{(l)}(t)$ denote the aggregated input-side and output-side messages; $AGG(\cdot)$ denotes asymmetric aggregators with distinct parameters; ${R}_i^{(0)}(t)={O}_i^{(0)}(t)=\mathbf{h}^{(0)}_{i}(t)$; $\sigma(\cdot)$ is the ReLU function. At each aggregation layer $l$, ${R}_j(t)$ and ${O}_j(t)$ are normalized by the maximum transaction amounts along the upstream and downstream directions at time $t$. This normalization is used to characterize the relative importance of interactions between firms in the supply chain.

Unlike traditional GNNs that process neighborhood information uniformly, this
mechanism explicitly differentiates between ``receiving'' and ``sending'' flows,
thereby capturing the structural asymmetries inherent in directed supply networks.
The receiving embedding ${R}_i(t)$ aggregates information about the production capacity and input structure of upstream firms, thereby reflecting the supply base of the focal firm. Conversely, the sending embedding ${O}_i(t)$ integrates information about the demand states of downstream firms, thereby capturing market absorption capacity. After $n$ layers, these two embeddings are fused through element-wise summation to obtain the comprehensive hidden state:
\begin{align}
    {Z}_i(t) = {R}_i^{(l)}(t) + {O}_i^{(l)}(t).
\end{align}

\begin{figure}
    \centering
    \includegraphics[width=0.8\linewidth,keepaspectratio]{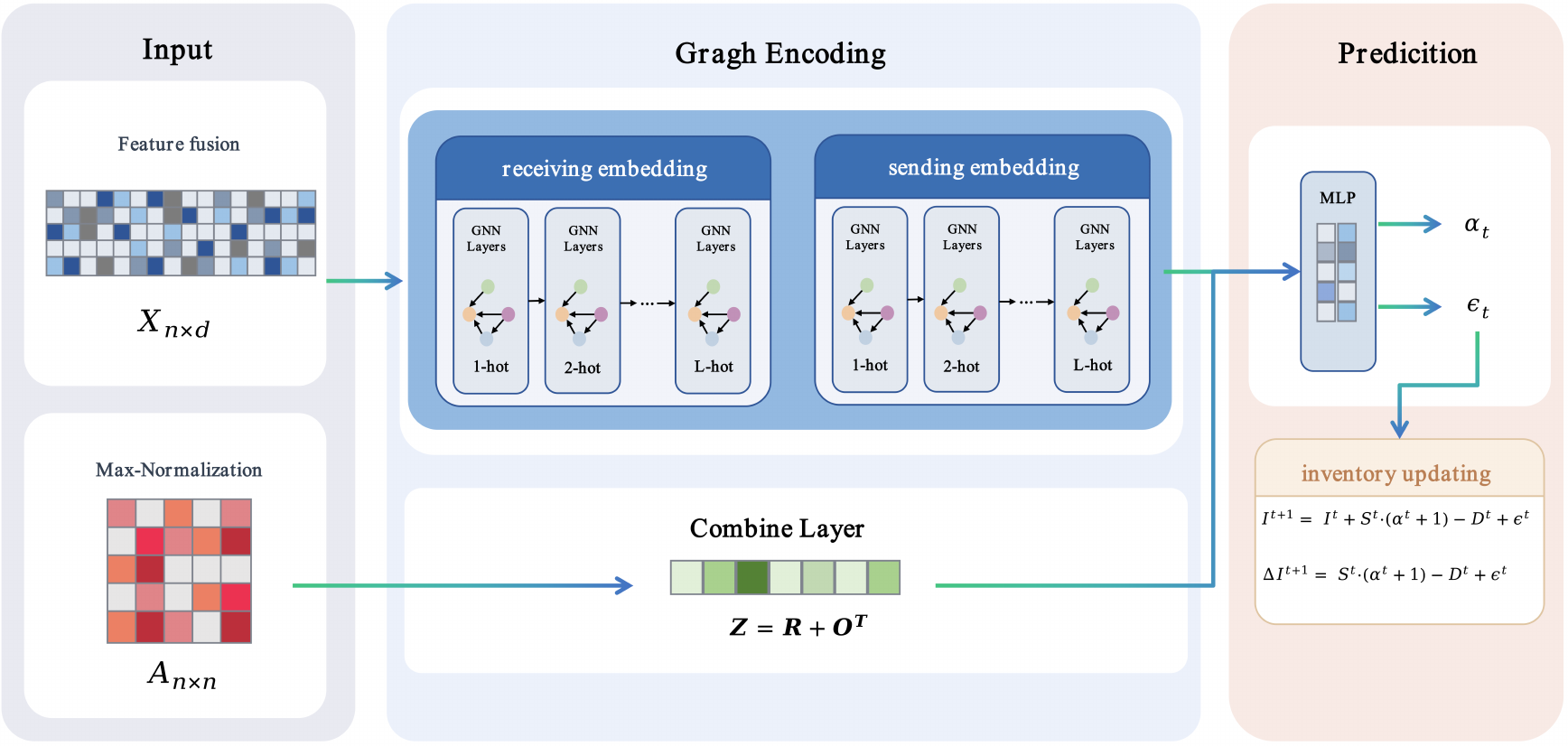}
    \caption{Architecture of the Directed Graph Neural Networks for Inventory Changes Prediction}
    \label{fig:gnn}
\end{figure}

We do not use ${Z}_i(t)$ directly to predict inventory changes. Instead, we first use ${Z}_i(t)$ to estimate two dynamic parameters, $\alpha_i(t)$ and $\epsilon_i(t)$. Needless to say, $\alpha_i(t)$ is an endogenous effective production coefficient learned by the model during inventory evolution, rather than an exogenously specified technological productivity parameter. $\epsilon_i(t)$ captures the unobserved dynamic adjustment component in the inventory system, rather than a conventional random noise term. Because internal variables such as production efficiency, capacity utilization, managerial capability, production loss, delivery delays, and inventory counting errors are generally unobservable for SMEs, we cannot directly measure $\alpha_i(t)$ and $\epsilon_i(t)$ using firm-level production data. We therefore rely on upstream and downstream structural information in the supply chain network to identify these two latent parameters.

We allow $\alpha_i(t)$ to depend jointly on ${R}_i(t)$ and ${O}_i(t)$ therefore uses supply-side and demand-side network information to identify the unobservable production adjustment coefficient.The productivity of a firm is not determined solely by its own characteristics, but is influenced by the structure of its supply chain relationships. \citet{SerpaKrishnan2018} show that productivity spillovers arise in vertical supply chain relationships and depend on supply chain structure. \citet{WangLiWuAnupindi2021} demonstrate that subtier supply network structure is an important source of focal firm risk. \citet{BimpikisCandoganEhsani2019} further show that network hierarchy, production costs, and disruption risks jointly shape firm profits. These studies suggest that upstream and downstream network structures contain information about resource access, demand response, risk exposure, and the operating environment of firms.

The same logic applies to $\epsilon_i(t)$. In the inventory updating equation, $\epsilon_i(t)$ absorbs inventory deviations that cannot be fully explained by explicit inflows, outflows, and the production function, including delivery instability, input loss, temporary inventory reallocation, order cancellations, counting errors, and inventory record inaccuracy. These deviations are not purely internal to the firm; they often arise from upstream supply frictions and downstream demand disturbances. \citet{chehrazi2025} shows that unobservable losses in inventory records are affected by demand and transactions, and \citet{FariasLiPeng2024} show that inventory record inaccuracy is common in real operations and difficult to identify using only longitudinal historical data. Thus, allowing $\epsilon_i(t)$ to depend on ${R}_i(t)$ and ${O}_i(t)$ enables the model to capture network-induced inventory adjustments that are difficult to observe directly.

In summary, we jointly predict $\hat{\alpha}_i(t)$ and $\hat{\epsilon}_i(t)$ by feeding ${Z}_i(t)$ into a Multi-Layer Perceptron (MLP), 
\begin{align}
    [{H}_{\alpha,i}(t), {H}_{\epsilon,i}(t)] &= \text{MLP}({Z}_i(t);\Theta), \\
    \hat{\alpha}_i(t) &= c_{\alpha} \tanh({H}_{\alpha,i}(t)), \\
    \hat{\epsilon}_i(t) &= \text{linear}({H}_{\epsilon,i}(t)),
\end{align}
where $c_{\alpha}$ is a scaling parameter controlling the range of production adjustments. 
Base on equation~\eqref{eq:effective_capacity} - \eqref{eq:inventory_change}, we calculate the inventory $\hat{I}_i(t)$ and inventory changes $\Delta \hat{I}_i(t)$ as follows:
\begin{align}
    \hat{Q}_i(t) &= S_i(t)(1+\hat{\alpha}_i(t)),\\
    \hat{I}_i(t) &= \hat{I}_i(t-1) + \hat{Q}_i(t) - D_i(t) + \hat{\epsilon}_i(t), \\
    \Delta \hat{I}_i(t) &= \hat{Q}_i(t) - D_i(t) + \hat{\epsilon}_i(t).
\end{align}

In our setting, a central challenge is that inventory data for SMEs are generally not observable in practice. To address this issue, we adopt an industry-based initialization strategy. Specifically, for firm $i$, initial inventory is obtained by scaling the average inventory level of listed firms in the same industry: 
\begin{equation}
I_i(0)=
\begin{cases}
I_i(0), & i \in \mathcal{V}^{listed},\\
\gamma_i \cdot \overline{I}_{s(i)}^{listed}(0), & i \in \mathcal{V}^{SMEs},
\end{cases}
\label{eq:initial_inventory}
\end{equation}
where $\overline{I}_{k}^{listed}(0)$ represents the average inventory of listed companies within industry $k$ to which firm $i$ belongs and $\gamma$ is the decay coefficient reflecting the difference in inventory levels between SMEs and the listed firms. Specifically, if a listed firm was not listed at time $t=0$, we set  $\gamma=0.5$. If the firm is SEMs, we set  $\gamma=0.1$.

Because our analysis focuses on the first difference $\Delta I_i(t)$, the influence of the initial inventory value is effectively removed,
\begin{equation}
I_i(t) = I_i(0) + \sum_{\tau=1}^{t} \Delta I_i(t).
\end{equation}

Finally, taking the first difference of the inventory level yields equation~\eqref{eq:inventory_change} , which is independent of the initial value.

\subsection{Loss Function Design}

To learn firm-level inventory changes in the absence of true inventory labels for SMEs, while ensuring that model predictions satisfy basic economic constraints and physical feasibility, we construct a hybrid loss function composed of three components. This loss function jointly constrains the learning process along three dimensions: predictive accuracy, nonnegativity, and parameter regularization.

For the subset of listed firms that publicly disclose financial statements, 
we use the Huber loss as the primary supervisory signal. 
Supply chain inventory and transaction data typically exhibit pronounced long-tail or power-law characteristics. 
For example, in product sales, a small number of highly popular items may account for most of total sales, 
while a large number of niche items have low individual sales volumes but remain collectively substantial. 
Such data therefore often contain a considerable number of outliers. 
The Huber loss provides a balance between mean squared error (MSE) and mean absolute error (MAE), 
preserving sensitivity to small errors while effectively limiting the excessive influence of extreme outliers during gradient updates. 
The supervised loss is defined as:
\begin{equation}
    \mathcal{L}_{sup} = \frac{1}{|\mathcal{V}_{listed}|} \sum_{i \in \mathcal{V}_{listed}} \text{Huber}(\hat{I}_i(t), I_i^{true}(t), \delta),
\end{equation}
where $\hat{I}_{i}(t)$ denotes the inventory value predicted by the model for firm $i$ at time $t$, 
$I_i^{true}(t)$ denotes the true inventory value for listed firms, and $\delta$ is the robustness threshold, 
which is set at $0.5$ in the experiments.

To prevent the model from generating physically meaningless negative inventory predictions during recursive updates, we introduce a penalty for violating the non-negative inventory constraint based on the ReLU function:
\begin{equation}
    \mathcal{L}_{zero} = \frac{1}{|\mathcal{V}|} \sum_{i \in \mathcal{V}} \text{ReLU}(-\hat{I}_i(t)).
\end{equation}

Building on this, we further include a standard parameter regularization term to control model complexity. 
The final optimization objective is defined as:
\begin{equation}\label{eq:overallLoss}
    \mathcal{L} = \mathcal{L}_{sup} + \lambda_1 \mathcal{L}_{zero} + \lambda_2 \|\mathbf{W}\|_2^2,
\end{equation}
where $\|\mathbf{W}\|_2^2$ is the $L_2$ norm of the parameters aimed at preventing model overfitting. $\mathbf{W}=\{\mathbf{W}_f, \mathbf{b}_f, \mathbf{W}_r^{(l)}, \mathbf{W}_o^{(l)}, \Theta\}$, 
$\lambda_1$ and $\lambda_2$ are hyperparameters for the non-negativity constraint and regularization.

\section{A Multi-Agent Econometric Framework for Validating Predicted Inventory Changes}
To validate the inventory predictions derived in the preceding section, we design an independent verification mechanism. Specifically, the framework deploys five agents, each associated with a distinct classical spatial econometric model, to evaluate the economic consistency of the predicted inventory changes.

The primary objective of this mechanism is to assess whether the predicted inventory changes are economically plausible by examining them across multiple dimensions, including dynamic persistence, causal direction, and network spillovers. Based on these analytical results, we infer the mechanisms underlying the predicted dynamics. Accordingly, this multi-dimensional approach provides a rigorous basis for evaluating the economic consistency of the predictions.

The core innovation of our approach lies in a comprehensive analytical workflow that directly evaluates the plausibility of the predicted inventory changes. On the input side, rather than relying exclusively on structured data, we use the predicted inventory changes as the primary input. By integrating these dynamics with the economic knowledge embedded in large language models, the framework guides the analytical process and helps align it with fundamental economic logic. On the output side, the framework goes beyond the mere reporting of numerical results from econometric models. Instead, each agent is required to generate interpretations with explicit economic meaning based on the computed results. Through this design, fragmented statistical outputs are transformed into unified, mechanism-based conclusions, thereby enabling a rigorous assessment of the economic plausibility of the inventory predictions.

\subsection{Theoretical Design of Our Multi-Agent Econometric Framework}
We design a multi-agent econometric framework to assess whether the inventory prediction results are internally consistent. Specifically, we analyze the same set of predicted inventory changes using a Spatial Panel Model\citep{wang2021} , a Dynamic Panel Model\citep{larson2015}, a Cross-Lagged Panel Model\citep{osadchiy2021}, a Network DID Model\citep{carvalho2021}, and a Supply-Demand Forecasting Model\citep{KOSCHAT2008}. If these econometric models yield consistent conclusions along key dimensions, we regard the corresponding inventory predictions as economically plausible. 

Existing studies have examined the relationship between inventory and supply-demand dynamics from a variety of econometric perspectives. Using a panel regression model, \citet{chiang2023} found that demand variation affects inventory dynamics. \citet{carvalho2021} leverage the Great East Japan Earthquake as a natural experiment and utilize a difference-in-differences (DID) model to investigate how supply chain shocks propagate across upstream and downstream partners, while \citet{baqaee2020} analyze the joint dynamics of supply and demand shocks within production networks. In parallel, a growing body of research has begun to examine the role of network effects in shaping inventory behavior and supply-demand relationships, arguing that supply chain network structure, firm centrality, and upstream-downstream linkages influence both inventory efficiency and shock propagation pathways \citep{agrawal2023, baqaee2024}. Building on this line of inquiry, \citet{wang2021} adopt a Spatial Durbin Model to analyze the inventory performance of manufacturing firms from a spatial spillover perspective, demonstrating a significant regional linkage effect in inventory fluctuations.. Extending these econometric foundations, we embed spatial effects into a supply-demand forecasting framework and incorporates network-weighted measures of supply, demand, inventory, and shocks. The econometric models supporting our analysis and their theoretical formulations are reported in Table \ref{tab:econometric_models}.

\begin{table}[htbp]
\caption{Summary of Econometric Models Embedded in the Multi-Agent Framework}
\centering
{\def\arraystretch{2.0} 
\begin{tabular}{p{0.22\textwidth} p{0.44\textwidth} p{0.28\textwidth}}
\hline
\hline 
\textbf{Analytical Module} & \textbf{Mathematical Formulation} & \textbf{Primary Analytical Objective} \\
\hline 

\textbf{1. Spatial Panel Model} 
& $Y_{kt} = \rho \sum_{m \neq k} W_{kmt} Y_{mt} + \beta X_{kt} + \mu_k + \varepsilon_{kt}$ 
& Identify spatial dependence in inventory changes, distinguishing between co-movement ($\rho > 0$) and hedging effects ($\rho < 0$) across interconnected industries. \\

\textbf{2. Dynamic Panel Model} 
& $Y_{kt} = \alpha Y_{k,t-1} + \beta_1 S_{kt} + \beta_2 D_{kt} + \mu_k + \varepsilon_{kt}$ 
& Evaluate inventory persistence ($\alpha$) and assess whether industries exhibit sustained destocking or accumulation dynamics. \\

\textbf{3. Cross-Lagged Panel Model} 
& $D_{kt} = \alpha_1 D_{k,t-1} + \beta_1 S_{k,t-1} + \mu_k + \varepsilon_{1,kt}$ \newline
  $S_{kt} = \alpha_2 S_{k,t-1} + \beta_2 D_{k,t-1} + \mu_k + \varepsilon_{2,kt}$ 
& Identify the dominant causal direction by comparing lagged cross-effects ($\beta_1$ and $\beta_2$), distinguishing supply-driven from demand-driven dynamics. \\

\textbf{4. Network DID Model} 
& $Y_{kt} = \beta \text{Shock}_{kt} + \gamma \sum_{m \neq k} W_{kmt} \text{Shock}_{mt} + \mu_k + \varepsilon_{kt}$ 
& Quantify both direct and network-mediated effects of exogenous shocks, capturing cascading propagation through the supply chain network. \\

\textbf{5. Supply-Demand Forecasting Model} 
& $S_{k,t+1} = \gamma_1 Y_{kt} + \lambda_1 S_{kt} + \theta_1 \text{Shock}_{t+1} + \mu_k + \varepsilon_{1,kt+1}$ \newline
  $D_{k,t+1} = \gamma_2 Y_{kt} + \lambda_2 D_{kt} + \theta_2 \text{Shock}_{t+1} + \mu_k + \varepsilon_{2,kt+1}$ 
& Forecast dynamic supply–demand interactions and assess whether inventory acts as a buffer or an amplifier under external shocks. \\

\hline
\hline 
\end{tabular}
\label{tab:econometric_models}
}
\end{table}

Within this framework, we encapsulate five econometric models into independent analytical agents. Each agent analyzes predicted inventory changes from a distinct economic perspective. For example, the Spatial Panel Model Agent examines whether predicted inventory changes are consistent with the network transmission logic of industrial chains. This agent takes inter-industry transactions ($S(t)$ and $D(t)$) and predicted inventory changes ($\Delta \hat I(t)$) as inputs and evaluates whether the inventory changes of an industry are affected by those of its upstream and downstream industries. Specifically, the agent identifies upstream supply sectors and downstream demand sectors based on industry chain relationships and estimates the direction and magnitude of neighboring industries' effects on the focal industry's inventory changes. A significantly positive effect indicates synchronized inventory adjustment across related industries and suggests the presence of network spillover effects. A significantly negative effect suggests inventory offsetting or inventory transfer among related industries.In addition, the Dynamic Panel Model Agent captures path dependence in inventory changes; the Cross-Lagged Panel Model Agent identifies the dominant direction between supply and demand; and the Network DID Model Agent and Supply-Demand Forecasting Model Agent examine external shocks and the buffering or amplifying role of inventory.

To prevent large language models from producing conclusions unsupported by the data, we impose strict constraints on the reasoning process of each agent. Specifically, each agent is allowed to conduct its analysis only on the basis of the computed results from the corresponding econometric model and may not engage in subjective inference detached from the data. At the same time, the conclusions of each agent must be explicitly grounded in the model output. Through this design, we ensure that the entire reasoning process remains anchored in verifiable quantitative evidence.

After obtaining the analytical results from the five agents, we further introduce an expert review agent to conduct a unified consistency check across conclusions generated by different econometric models. This mechanism is primarily used to identify potential logical conflicts across agents, especially inconsistencies on key issues such as the causal direction between supply and demand, and to revise and integrate these conclusions accordingly. Through this process, we improve the overall consistency and reliability of the final analytical results.

Finally, the multi-agent system generates a comprehensive evaluation report for each industry. This report includes not only the numerical results of each econometric model but also an integrated synthesis of conclusions across different analytical dimensions. It is intended to systematically characterize the inventory dynamics of each industry, including network spillover effects, dynamic adjustment capability, dominant driving mechanisms, and responses to external shocks. Based on these results, we can further assess whether the inventory predictions are broadly consistent with economic regularities, thereby providing an effective validation of the prediction results.

\begin{figure}
    \centering
    \includegraphics[width=0.91\linewidth,keepaspectratio]{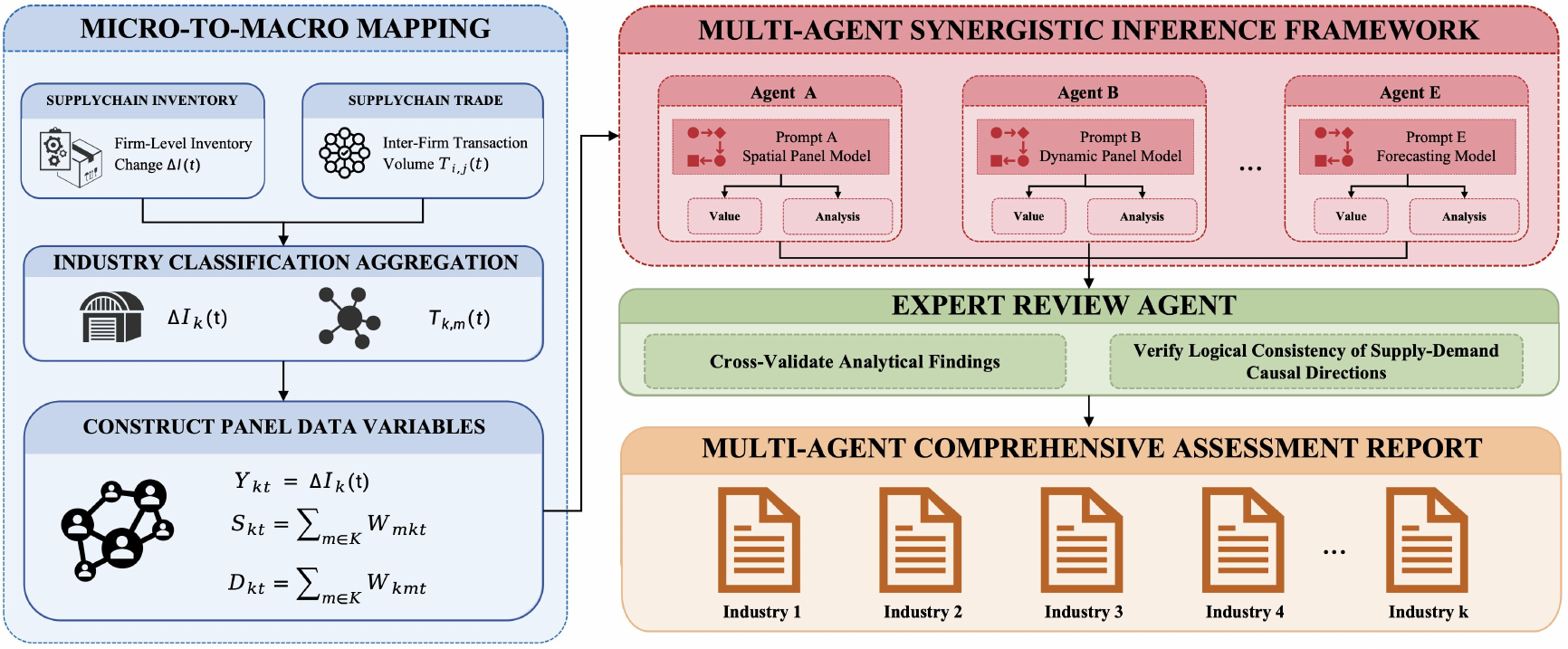}
    \caption{The Multi-Agent Econometric Framework for Validating Predicted Inventory Changes}
    \label{fig:agent}
\end{figure}

\subsection{Implementation Workflow of Our Multi-Agent Econometric Framework}
The overall workflow of the proposed multi-agent validation framework is illustrated in the Figure~\ref{fig:agent}. For each industry, we use industrial-chain network data as the input and provide the data separately to five agents. Each agent loads one econometric model and analyzes the predicted inventory changes from a clearly defined economic perspective. These perspectives include spatial spillovers, dynamic persistence, causal direction identification, shock transmission, and supply-demand forecasting.

\begin{table}[htbp]
\caption{Agent Skill and Output}
\centering
{\def\arraystretch{1.5}
\begin{tabular}{p{0.46\textwidth} p{0.48\textwidth}} 
\hline
\hline
\textbf{Agent Skill} & \textbf{Output} \\
\hline

\textbf{Role:} Expert in Chinese industries and supply chains. \newline
\textbf{Task Description:} Based on the methodology of [Model], infer the causal characteristics of \{industry\_name\}. \newline
\textbf{1. Methodology and Core Formulas} \newline
- Core formulation \newline
- Primary problems identified and addressed \newline
- Implementation pathway \newline
\textbf{2. Input Data Description} \newline
\textbf{3. Analysis and Output Format Requirements} \newline
\textbf{Step 1: Computation and Analysis} \newline
\texttt{\{"Key computational processes and result analysis."\}} \newline
\textbf{Step 2: JSON Output} \newline
\texttt{\{ "shock\_event": "Name of the introduced shock", \newline
"beta": "Numeric value, num", \newline
"gamma": "...", \newline
"feature1": "Characteristic reflected by the value, str", \newline
"feature2": "...", \newline
"reason": "Reasoning, str", \newline
"summary": "Textual analysis results, str" \}} 
&
\textbf{Industry\_Name:} \newline 
\texttt{"GICS Industry Name",} \newline
\textbf{Expert\_Models\_Details: \{} \newline
\texttt{"Agent 1: model 1": \newline
"Computational process..., numerical values..., \newline
characteristics..., reasoning analysis....", \newline
"Agent 2: model 2": "...", \newline
"Agent 3: model 3": "...", \newline
"Agent 4: model 4": "...", \newline
"Agent 5: model 5": "..."\},} \newline
\textbf{Expert\_Correction\_Process: \{} \newline
\texttt{"logic\_check": "...", \newline
"contradictions\_found": "...", \newline
"correction\_reasoning": "...", \newline
"final\_synthesis": "...", \newline
"core\_dimension\_1": "Demand-driven", \newline
"core\_dimension\_2": "Proactive destocking" \}} \\
\hline
\hline
\end{tabular}
\label{tab:prompt_output_example}
}
\end{table}

As shown in Table~\ref{tab:prompt_output_example}, each agent independently loads one econometric model and conducts a structured analysis of the input data during execution. More specifically, the agent first constructs the computational procedure according to the definition of the model and then performs the numerical calculations strictly based on the prespecified formulas. It subsequently generates the corresponding economic interpretation on the basis of the computed results.

To prevent large language models from generating conclusions unsupported by the data during reasoning, we impose strict constraints on the execution process of each agent. Specifically, each agent must first present the complete calculation process and only then provide its conclusion. In addition, the conclusion produced by each agent must remain consistent with the computed results of the corresponding econometric model. Through this design, we ensure that the entire reasoning process remains grounded in verifiable quantitative evidence.

After obtaining the independent analytical results from the five agents, we further introduce an expert validation mechanism to examine the conclusions across agents. This mechanism focuses on identifying potential logical conflicts among agents, such as inconsistencies between inferred causal directions and observed characteristics of inventory dynamics. When such conflicts are detected, the system reexamines and adjusts the relevant conclusions on the basis of the fundamental logic of relationship between supply and demand, thereby producing an integrated assessment that is logically coherent.

Finally, the multi-agent system generates a structured evaluation output for each industry. This output preserves both the calculation process and the original conclusions of each agent, while also integrating key industry characteristics, such as the dominant drivers of inventory changes and the corresponding stage of the inventory cycle. Based on these results, we can systematically assess whether the predicted inventory changes conform to basic economic regularities, thereby enabling an effective validation of the prediction results.

\subsection{Inventory Cycle Inference of Industry}
We map the firm-level supply chain results onto the broader industrial network. We conduct inventory-cycle inference at the industry level rather than at the firm level because firm-level inventory fluctuations are often affected by firm-specific operational decisions, reporting errors, and temporary transaction shocks. By contrast, industry-level inventory changes more accurately capture systematic supply-demand mismatches, which constitute the central focus of our analysis. Given the idiosyncratic nature of inventory in certain sectors, we intentionally exclude service industries, such as financial services and real estate, during the aggregation process. The final set of valid industries is defined as $\mathcal{K}'$ ($|\mathcal{K}'|= 58$). For each industry $k \in \mathcal{K}'$, the industry-level inventory variation, $\Delta I_k(t)$, is calculated by aggregating the contemporaneous inventory variations of all firms within that classification:
\begin{equation}
    \Delta I_k(t) = \sum_{i \in \mathcal{V}_k} \Delta I_i(t),
\end{equation}
where $\mathcal{V}_k \subset \mathcal{V}$ denotes the subset of firm nodes belonging to industry $k$, and $\Delta I_i(t)$ represents the predicted inventory changes of firm $i$ at time $t$.

Following a similar logic, we map firm-level transaction records onto the industrial network to establish input-output linkages across different sectors. Let $T_{ij}(t)$ denote the transaction volume from supplier $i$ to customer $j$. The aggregate transaction flow from industry $k$ to industry $m$ is defined as:
\begin{equation}
    W_{k \rightarrow m}(t) = \sum_{i \in \mathcal{V}_k} \sum_{j \in \mathcal{V}_m} T_{ij}(t)
\end{equation}

From the perspective of inventory-cycle theory, inventory changes provide a more natural lens through which to characterize supply-demand mismatch. Short-term economic fluctuations often arise from information delays, expectation errors, and sluggish production adjustment. Changes in demand typically occur before supply has fully adjusted, and inventory is precisely the state variable that absorbs this temporal gap. When supply persistently exceeds demand, inventory accumulates passively; when demand exceeds supply capacity, inventory is depleted; and when firms actively adjust production and procurement based on expectations, inventory enters either an active replenishment phase or an active destocking phase. Inventory changes therefore do not merely reflect the operating outcomes of firms; they also capture the intertemporal accumulation of supply-demand mismatch. Compared with one-period transaction flows, inventory changes more effectively reflect persistent divergence between supply and demand.

Based on the above theoretical framework, we define two core output dimensions for the multi-agent inference results: industry driving type and inventory cycle stage. The former is designed to identify the dominant force shaping inventory changes, namely whether changes in inventories are primarily driven by demand-side factors, supply-side factors, or the joint interaction of both. The latter is intended to characterize the current economic state of inventory changes within the inventory cycle framework, distinguishing among active inventory replenishment, passive inventory accumulation, active destocking, and passive destocking.

More specifically, if inventory changes of a industry are driven primarily by changes in downstream demand, we classify it as \emph{demand-driven}. If inventory changes
arise mainly from changes in upstream supply, production capability, or procurement input, we classify it as \emph{supply-driven}. If both supply-side and demand-side factors play significant roles, we classify it as \emph{bidirectionally driven}. For inventory cycle classification, if firms expand production or procurement in anticipation of improving demand and this leads to rising inventory, the corresponding phase is \emph{active replenishment}. If demand declines while supply adjustment lags, causing inventory to rise passively, the phase is \emph{passive accumulation}. If firms actively compress production or procurement in order to reduce inventory pressure, the phase is \emph{active destocking}. If demand recovers or sales improve faster than supply can be replenished, causing inventory to decline passively, the phase is \emph{passive destocking}.

\section{Experiments}
\subsection{Experimental Setup}

The ChinaScope dataset used in this study covers supply-chain transaction records from 2015 to 2023. 
We construct the supply chain graph by treating firms as nodes, interfirm transaction relationships as directed edges, 
and transaction values as edge weights. All firms are classified according to the Global Industry Classification Standard (GICS), 
jointly introduced by Standard \& Poor’s and Morgan Stanley in August 1999. All firms in our sample are mapped into 139 industry categories within this classification framework.

For data partitioning, we adopt a rolling time-window strategy to evaluate the model’s ability to generalize to future periods. Specifically, the training set covers data from December 2015 to December 2020, while the test set consists of data from June 2021 to June 2023.  During model training, nodes belonging to service industries such as finance and real estate are masked and excluded from the loss-function computation so as to prevent structural bias from interfering with model learning.

In terms of model architecture, the node embedding dimensions are configured as follows: firm ID is set to 32 dimensions, industry features to 16 dimensions, listing status to 4 dimensions, and seasonal features to 4 dimensions. Our model employs the asymmetric GNN architecture described in Section ~\ref{sec:DGNN}. We use the Adam optimizer coupled with a Cosine Annealing Scheduler, setting the minimum learning rate to $1 \times 10^{-4}$. To address potential outliers in the inventory data, we adopt the Huber loss function ($\delta=0.5$) as the primary optimization objective. In addition, we select the non-negativity penalty parameters $\lambda_1=0.2$ and  $\lambda_1=1 \times 10^{-5}$ (as shown in \eqref{eq:overallLoss}) with grid searching, thereby constraining the model outputs to be economically reasonable and improving the robustness of model training. To prevent overfitting, a Dropout rate of 0.2 is applied to both the feature fusion layer and the GNN layers. Model performance is evaluated using three metrics: Root Mean Square Error (RMSE), Mean Absolute Error (MAE), and Weighted Absolute Percentage Error (WAPE), which measure prediction error and stability from different perspectives.

\subsection{Results of Inventory Prediction}

\begin{table}[htbp]
\caption{Evaluation of Inventory Prediction Results}
\centering
{\def\arraystretch{1.5}
\begin{tabular*}{0.7\textwidth}{@{\extracolsep{\fill}}l c c c c}
\hline
\hline
\textbf{Date} & \textbf{Sample Size} & \textbf{RMSE} & \textbf{MAE} & \textbf{WAPE} \\
\hline
2021/06/30 & 3,916 & 24.2797 & 4.8392 & 0.2327 \\
2021/12/31 & 4,077 & 26.7815 & 6.2834 & 0.3004 \\
2022/06/30 & 4,124 & 29.8125 & 4.8030 & 0.2116 \\
2022/12/31 & 4,235 & 36.3061 & 7.5795 & 0.3481 \\
2023/06/30 & 4,264 & 27.6519 & 4.7379 & 0.2160 \\
\hline
\hline
\end{tabular*}
\label{tab:prediction_results}
}
\end{table}

Table \ref{tab:prediction_results} shows the model's inventory prediction performance across various time nodes during the period from June 2021 to June 2023. As the number of evaluated nodes gradually increases, the model maintains a stable predictive performance despite the growing network size, thereby demonstrating strong scalability and robustness.

The MAE of our model remains consistently low, ranging from 4.7379 to 7.5795 with only narrow fluctuations. This pattern indicates stable prediction accuracy. Similarly, WAPE remains stable, fluctuating between 0.2116 and 0.3481, which suggests that the model effectively controls relative errors across different time periods. Although the RMSE exhibits slight increases at specific time points, the overall variation is limited, indicating that the model is robust to localized fluctuations or anomalous shocks. Overall, the model exhibits strong stability and predictive capability across different periods, demonstrating its effectiveness in capturing long-term inventory changes and establishing it as a reliable baseline model for supply chain inventory forecasting.
\begin{figure}[!htp]
    \centering
    \subfigure[Dimension RMSE]{
        \label{fig:dim_rmse}
        \includegraphics[width=0.23\textwidth]{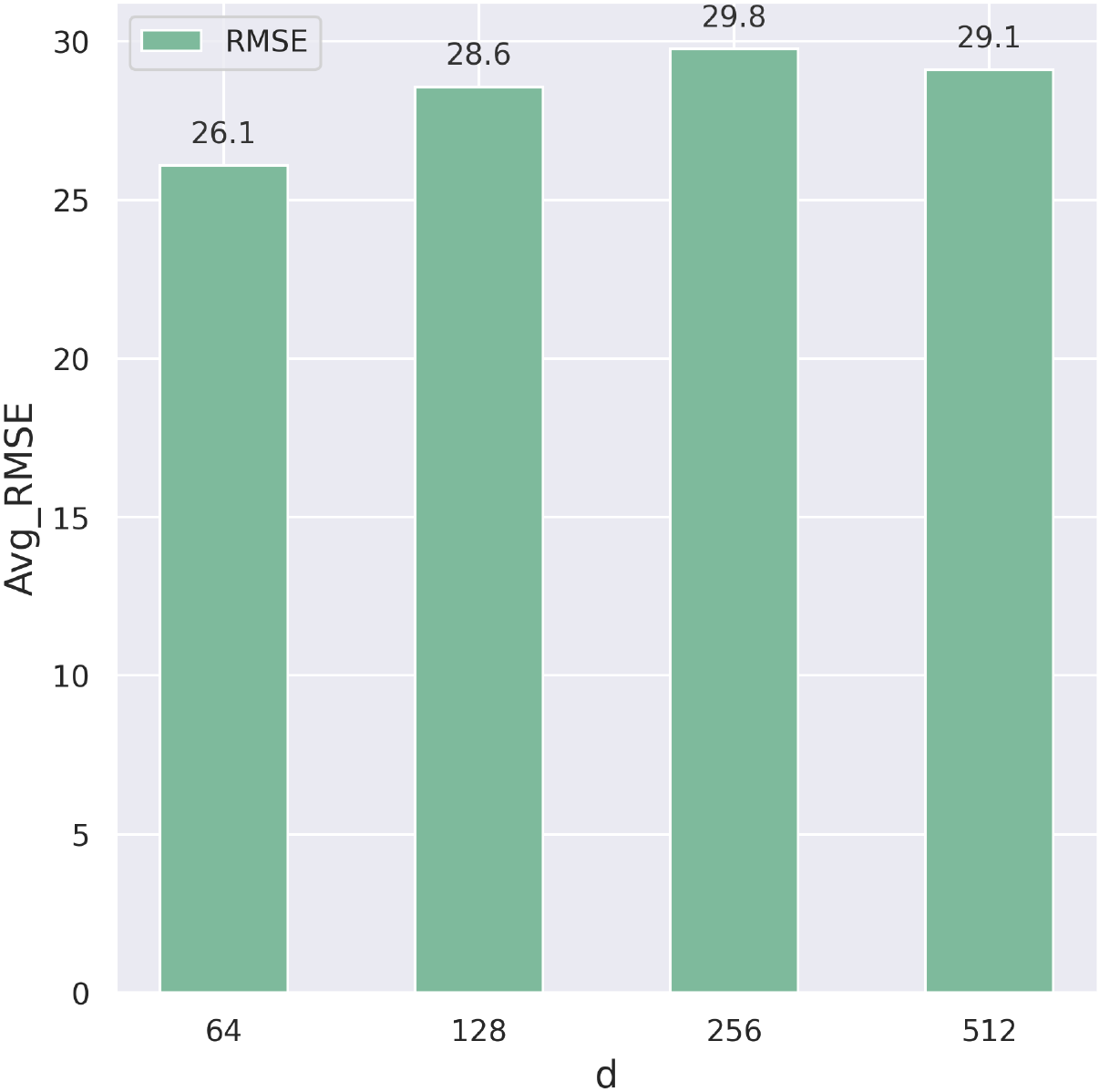}
    }
    \hfill
    \subfigure[Dimension MAE]{
        \label{fig:dim_mae}
        \includegraphics[width=0.23\textwidth]{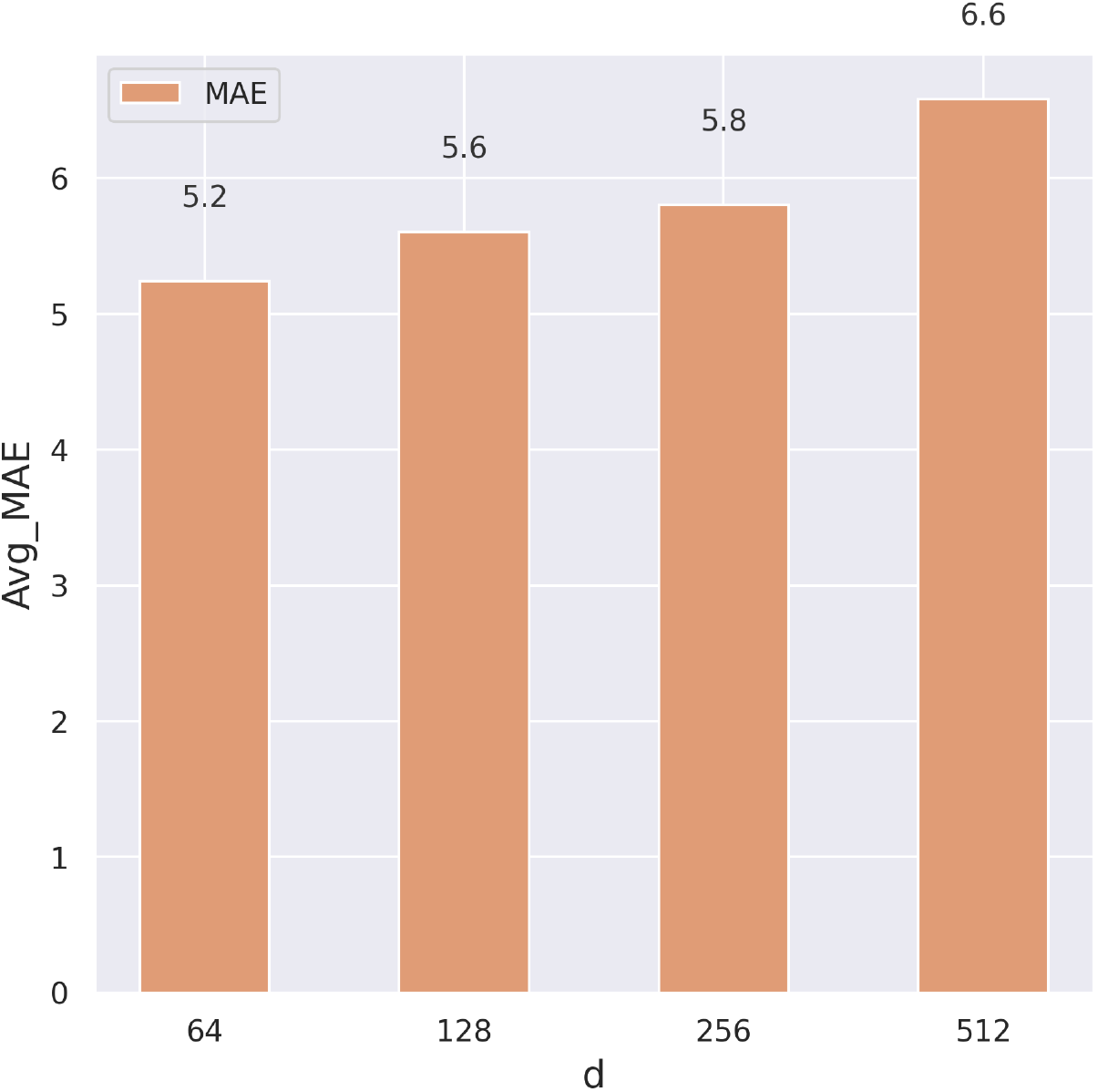}
    }
    \hfill 
    \subfigure[Layers RMSE]{
        \label{fig:layers_rmse}
        \includegraphics[width=0.23\textwidth]{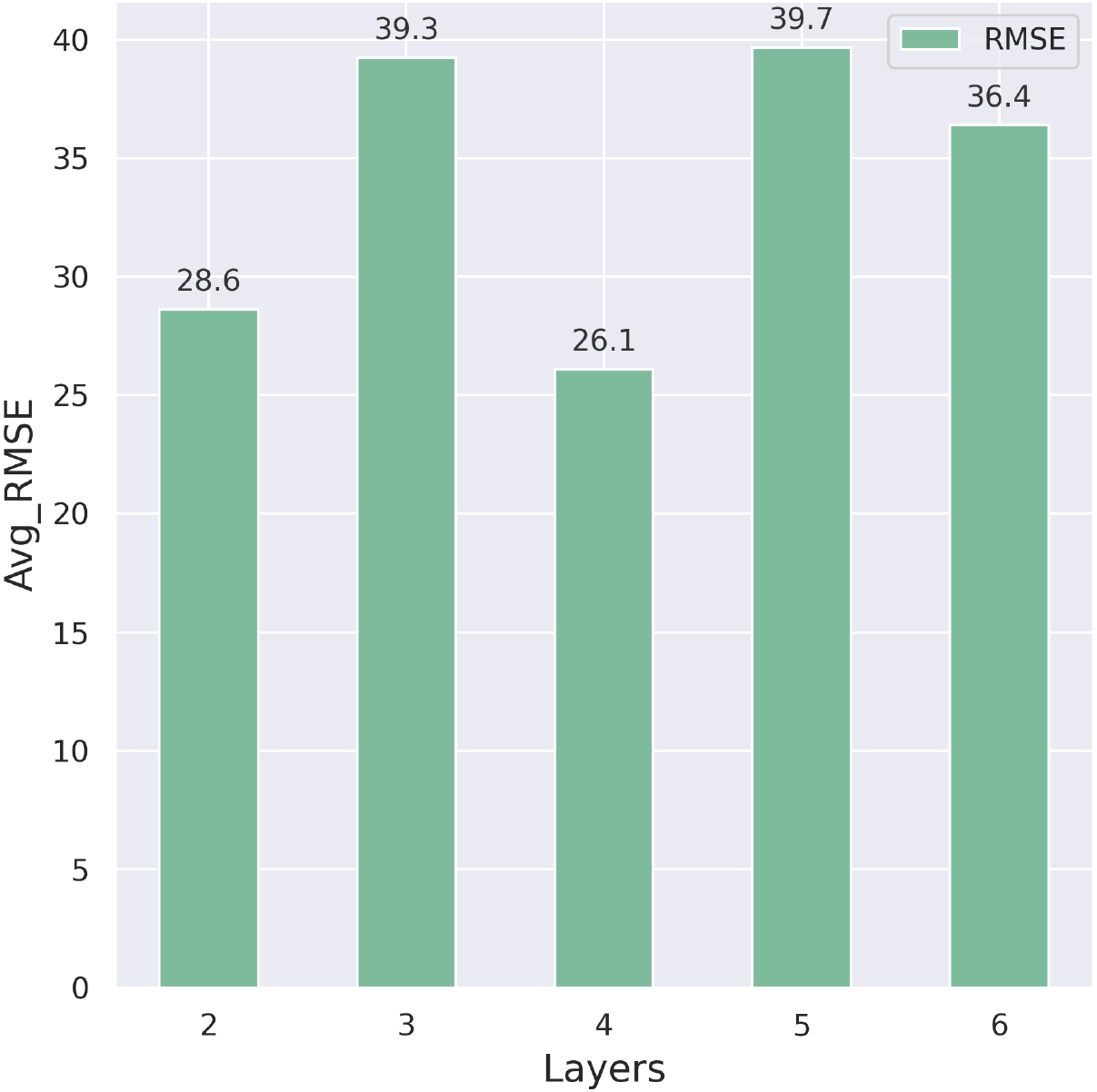}
    }
    \hfill
    \subfigure[Layers MAE]{
        \label{fig:layers_mae}
        \includegraphics[width=0.23\textwidth]{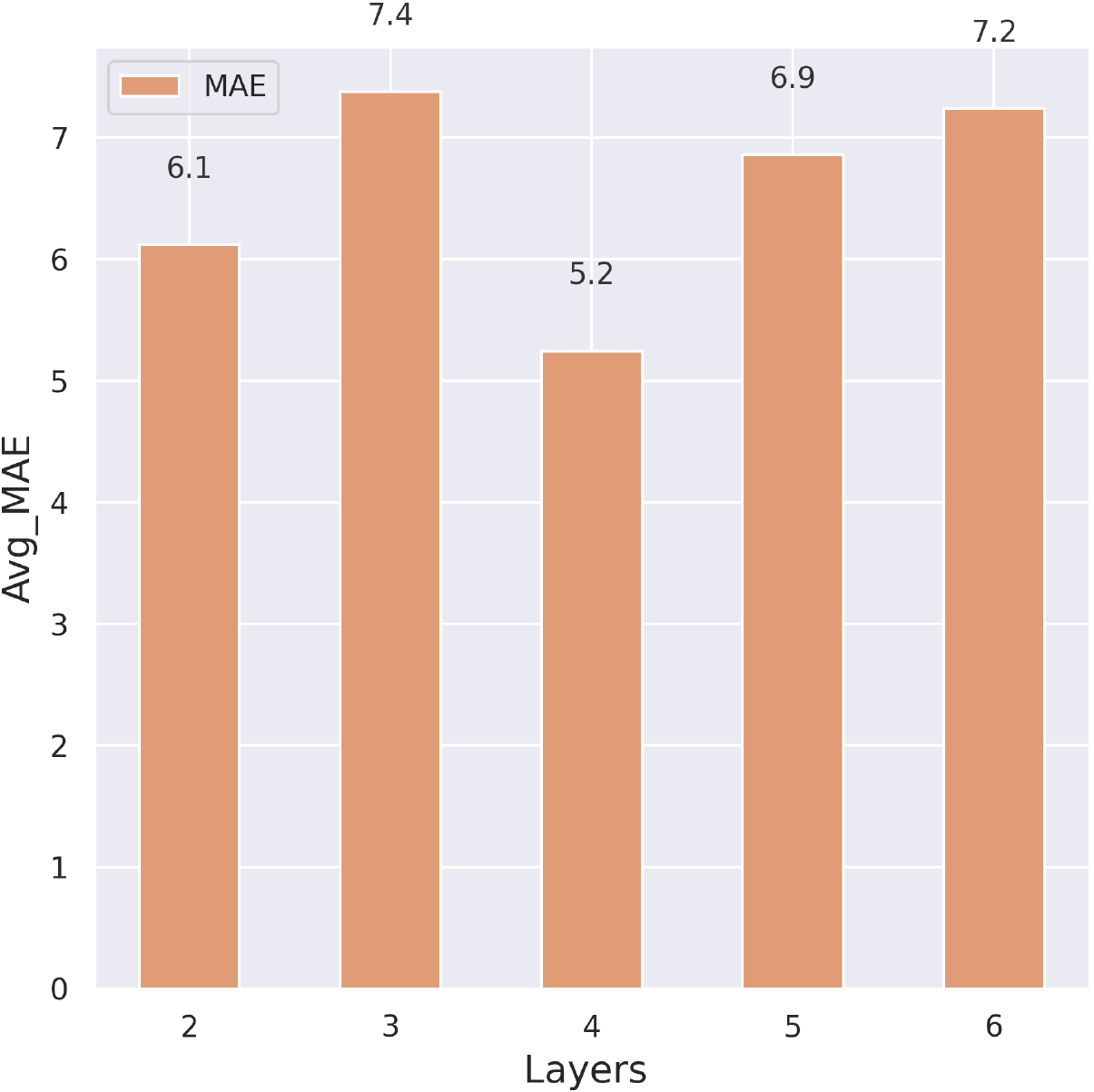}
    }
    
    \caption{Sensitivity Analysis of Embedding Dimension and Network Depth}
    \label{fig:all_sensitivity_analysis}
\end{figure}

In neural network architectures, we have found that feature dimension and network depth have a significant impact on the model’s predictive performance. As shown in Figure \ref{fig:all_sensitivity_analysis}\subref{fig:dim_rmse} - \ref{fig:all_sensitivity_analysis}\subref{fig:dim_mae}, 
the model achieves its best performance at 64 dimensions; further increases in dimensionality lead to higher errors, likely because of overfitting. As shown in Figure \ref{fig:all_sensitivity_analysis}\subref{fig:layers_rmse} - \ref{fig:all_sensitivity_analysis}\subref{fig:layers_mae}, the four layers model delivers the best predictive performance, whereas deeper architectures reduce performance, likely because of oversmoothing. 

\begin{figure}[!htp]
    \centering
    \subfigure[Param($\lambda_1$) RMSE]{
        \label{fig:Hyperparameter1_rmse}
        \includegraphics[width=0.23\textwidth]{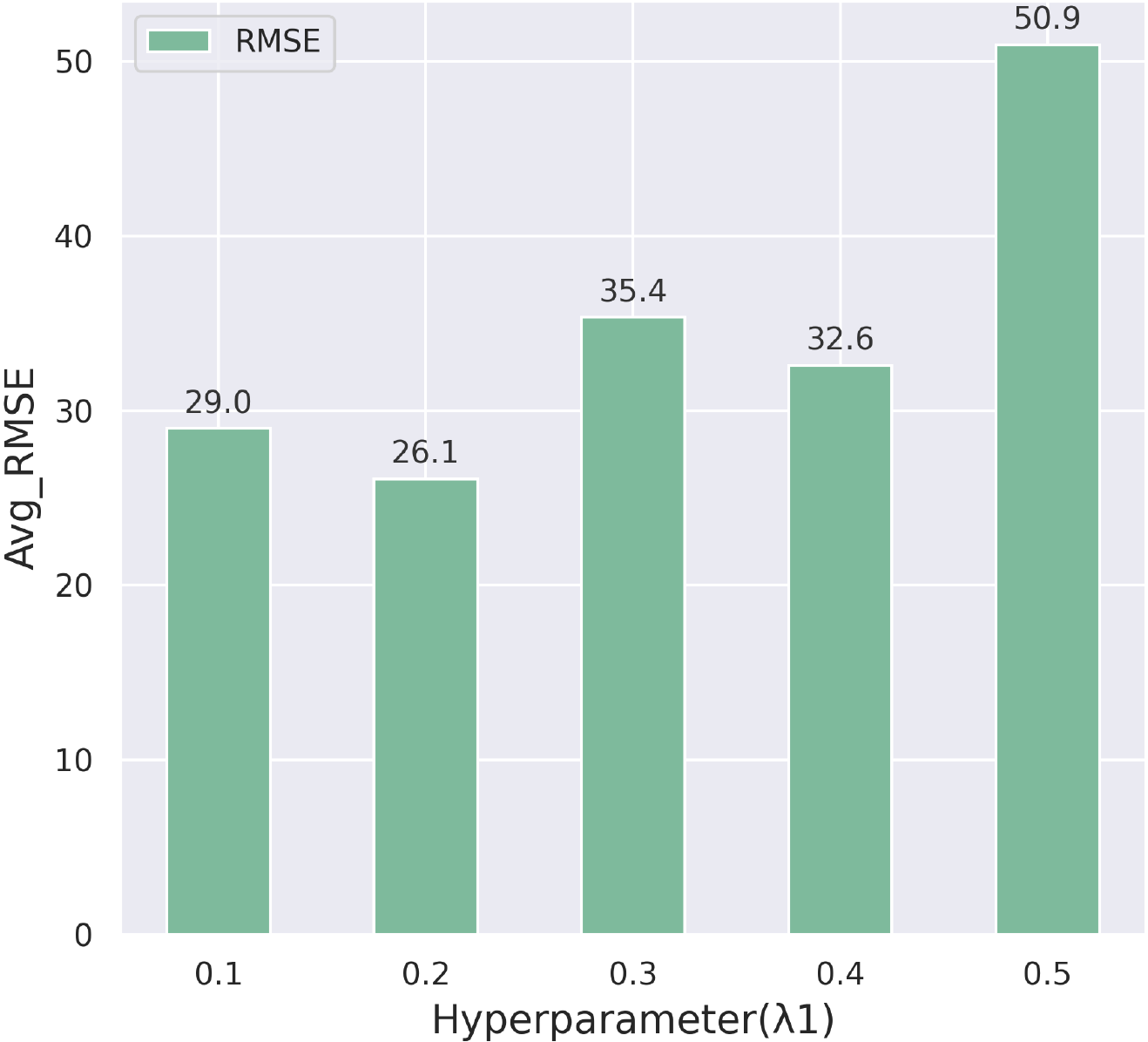}
    }
    \hfill
    \subfigure[Param($\lambda_1$) MAE]{
        \label{fig:Hyperparameter1_mae}
        \includegraphics[width=0.23\textwidth]{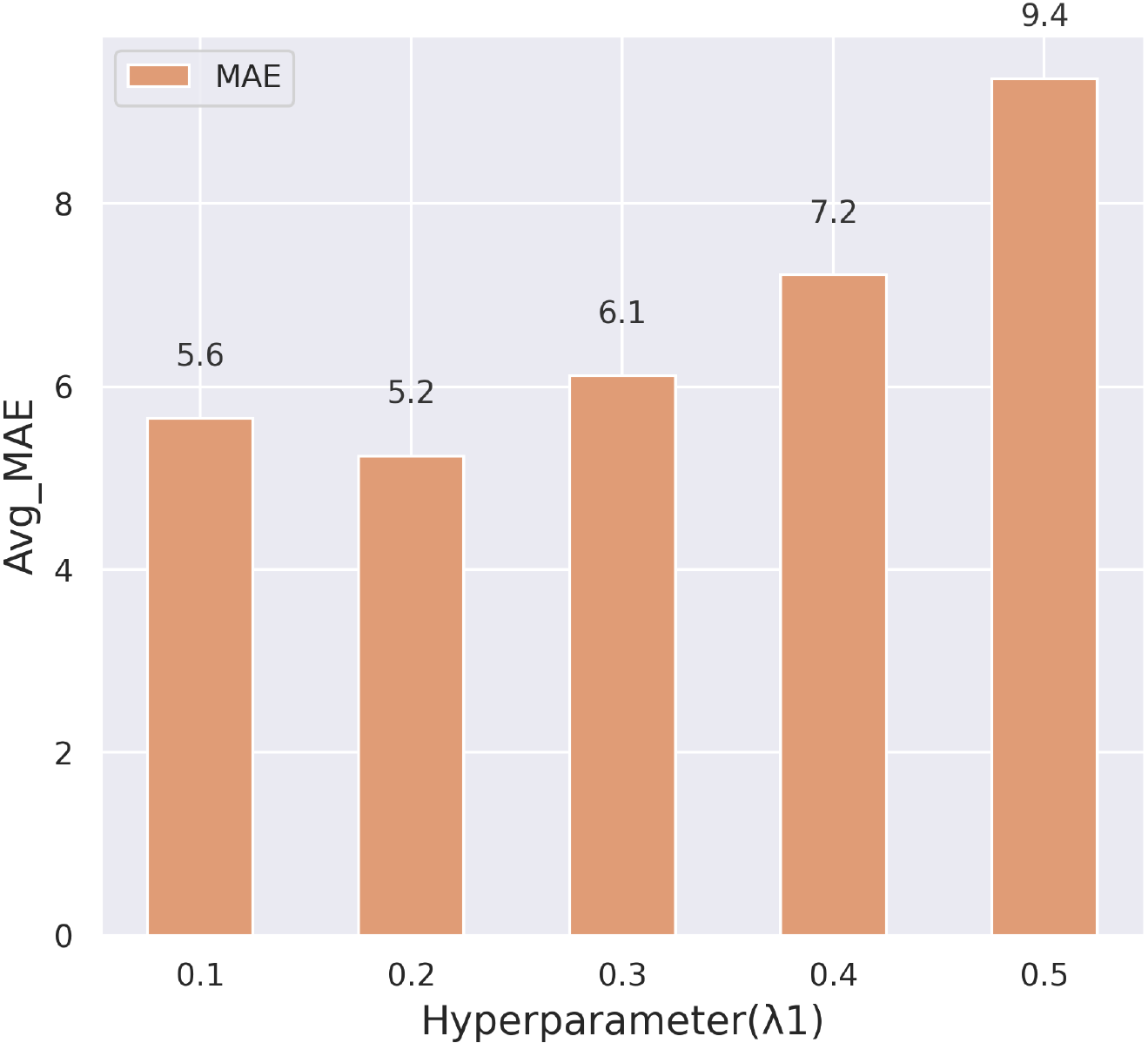}
    }
    \hfill
    \subfigure[Param($\lambda_2$) RMSE]{
        \label{fig:Hyperparameter2_rmse}
        \includegraphics[width=0.23\textwidth]{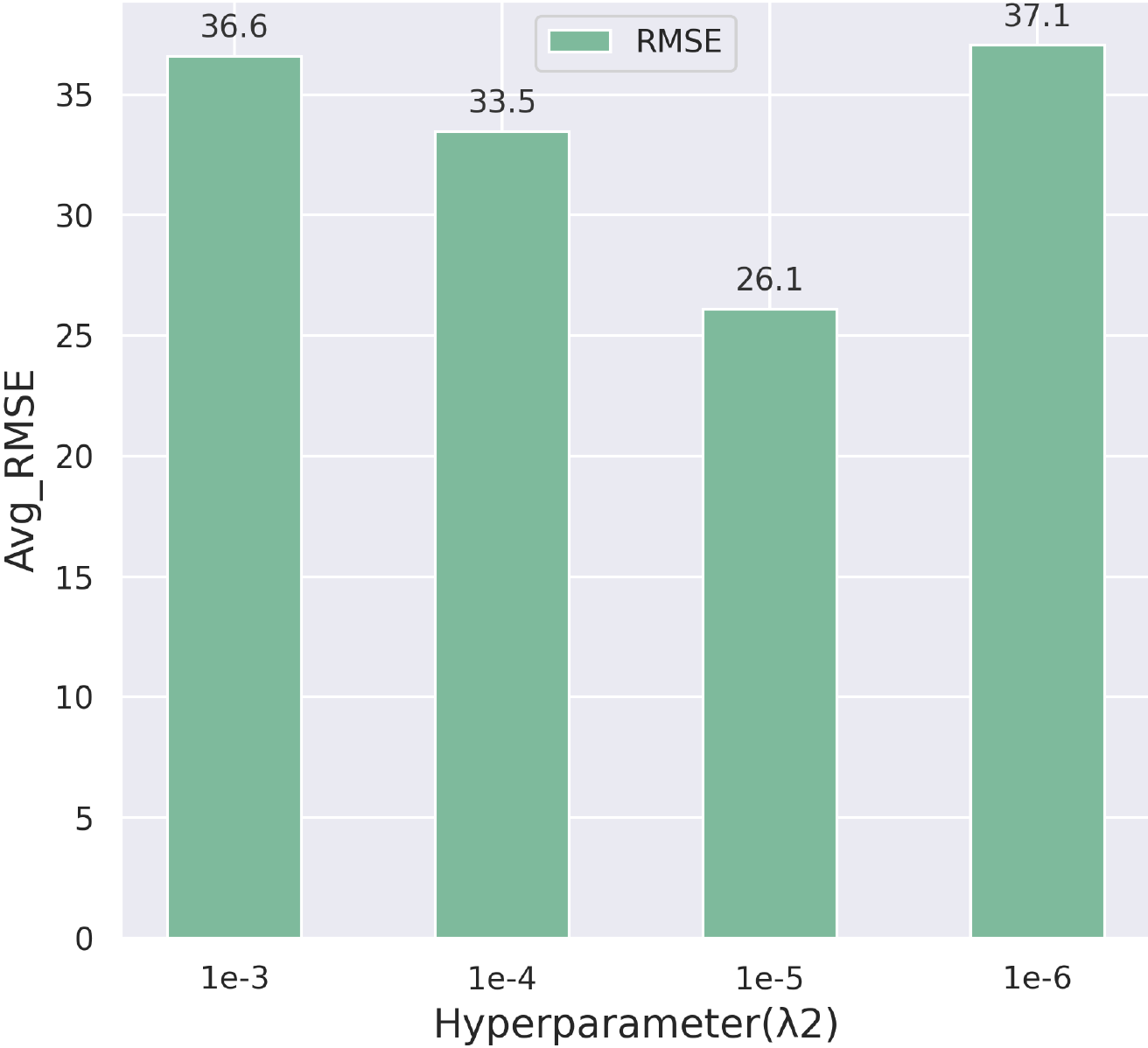}
    }
    \hfill
    \subfigure[Param($\lambda_2$) MAE]{
        \label{fig:Hyperparameter2_mae}
        \includegraphics[width=0.23\textwidth]{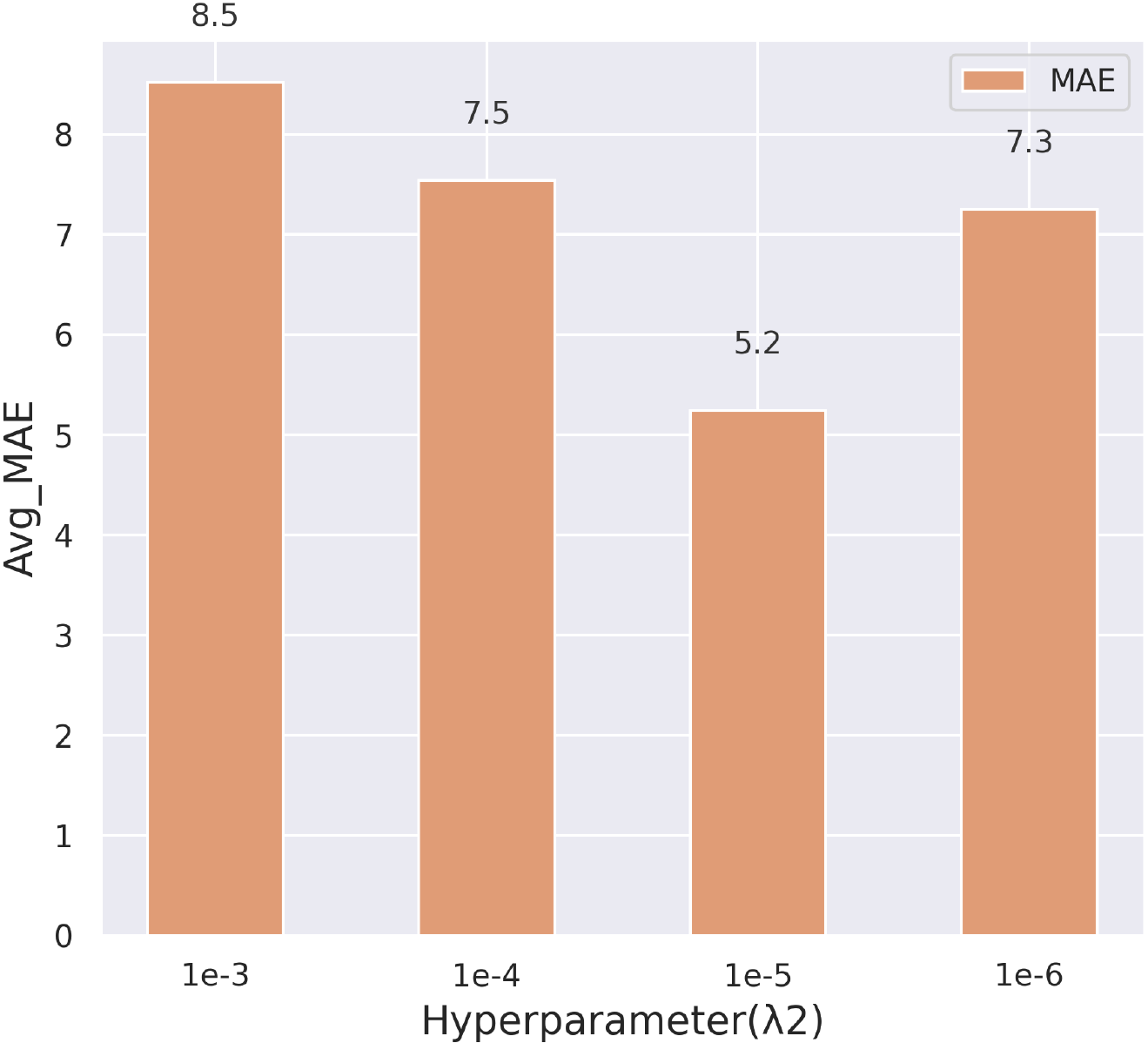}
    }
    
    \caption{Sensitivity Analysis of Hyperparameters $\lambda_1$ and $\lambda_2$}
    \label{fig:sensitivity_analysis_2}
\end{figure}

As shown in Figure \ref{fig:sensitivity_analysis_2}\subref{fig:Hyperparameter1_rmse} - \ref{fig:sensitivity_analysis_2}\subref{fig:Hyperparameter1_mae}, 
we find that setting the weight of the non-negative constraint parameter($\lambda_1$) to 0.2 effectively controls the error, whereas a higher value significantly reduces the model’s performance. This pattern indicates that overly restrictive constraints may limit the expressive capacity of the model.
For the regularization coefficient $\lambda_2$ ( Figure~\ref{fig:sensitivity_analysis_2}\subref{fig:Hyperparameter2_rmse} - \ref{fig:sensitivity_analysis_2}\subref{fig:Hyperparameter2_mae}), the model performs best when $\lambda_2 = 10^{-5}$. An excessively large weight decay may restrict the expressive capacity of the model, while an excessively small weight decay may reduce its generalization ability. 
Overall, the configuration of $\text{dim}=64$, $\text{layers}=4$, $\lambda_1=0.1$, and $\lambda_2 = 10^{-5}$ strikes a favorable balance between prediction accuracy and stability.

In summary, the experiment results demonstrate that our GNN model possesses excellent reliability, accuracy, and stability in the task of predicting long-term inventory changes. Therefore it can be served as an effective baseline framework for supply chain inventory prediction. 

\subsection{Results of Multi-Agent Consensus Validation}

To further examine the stability of the multi-agent analysis results, 
we ran the same econometric agent framework on three large language models: 
Claude-Sonnet-4.5, GPT-5.4, and Kimi-K2.5. 
Then, we evaluated the consistency of their output conclusions along two core dimensions. 
The judgments produced by the three LLMs were classified into three categories according to their consistency: 
``3/3 Full Agreement''``2/3 Majority Agreement'' and ``Disagreement''.

\begin{figure}
    \centering
    \includegraphics[width=1.0\linewidth]{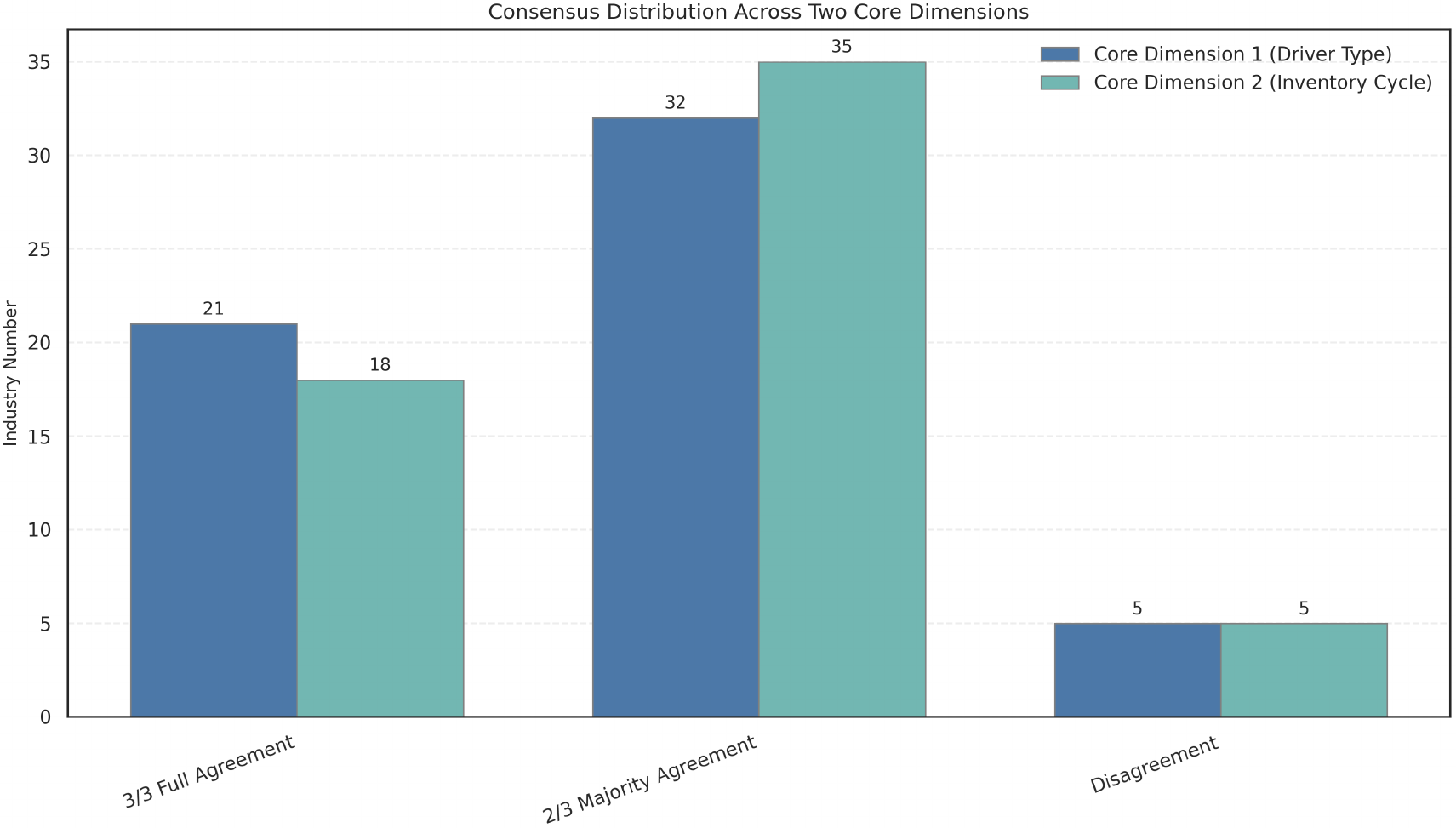}
    \caption{Consistency Between Industry Driving Types and Inventory Cycles}
    \label{fig:agent results}
\end{figure}

As shown in Figure~\ref{fig:agent results}, different large language models exhibit a high overall degree of agreement across the two core dimensions. Along the driving type dimension, 21 industries show full agreement across all three models and 32 industries show majority agreement at the 2/3 level, whereas only 5 industries exhibit substantial disagreement. This pattern suggests that, for the vast majority of industries, different models reach broadly consistent judgments about the dominant drivers of inventory changes. Agreement is also concentrated along the inventory cycle dimension: 18 industries exhibit full agreement and 35 industries exhibit majority agreement, whereas 5 industries show substantial disagreement. Overall, across both dimensions, industries achieving at least majority agreement account for a clear majority, indicating that the multi-agent framework yields relatively stable conclusions across models.

Although we apply the same econometric framework and expert-correction procedure to all large language models, some low-consensus cases still remain. Even under the same analytical setting, different large language models may interpret inventory dynamics in the same industry differently because of variation in training corpora, coverage of industry knowledge, and preferences in economic interpretation. In industries with strong domain specificity, explanations of inventory dynamics often depend on background knowledge about regulatory institutions, technology cycles, energy prices, capital expenditures, public-utility characteristics, or strategic reserves. Because models differ in how they acquire and use such knowledge, disagreements may still arise over the dominant driving forces or the stage of the inventory cycle, even after expert correction. Low-consensus results therefore do not imply that the predictions are invalid; rather, they indicate that the economic interpretation of these industries is more sensitive to the prior knowledge of models. By contrast, the large number of high-consensus results suggests that the corresponding conclusions are more robust across models.

Overall, our econometric agent framework demonstrates strong robustness across multiple large language models. When different models reach similar judgments based on the same econometric results, this suggests that the conclusions are not incidental products of any single model, but instead possess substantial cross-model robustness. Because all models in our framework reason on the basis of the same econometric specifications, the consistency of their conclusions reflects the stability of large language models in generating econometric interpretations. Put differently, multi-model consistency testing provides additional evidence supporting the economic interpretation of the predicted inventory changes.

\subsection{Case Analysis of Representative Industry Chains}
Building on the outputs of the econometric agents, we further conduct visual analyses of local industry chain subgraphs to examine predictive performance of our model in specific scenarios. By inspecting representative cases, we are able to show more intuitively both the situations in which multiple models reach consistent conclusions and those in which disagreement remains, thereby providing supplementary evidence for the consistency-based validation of the prediction results.

\begin{figure}
    \centering
    \includegraphics[width=1.0\linewidth]{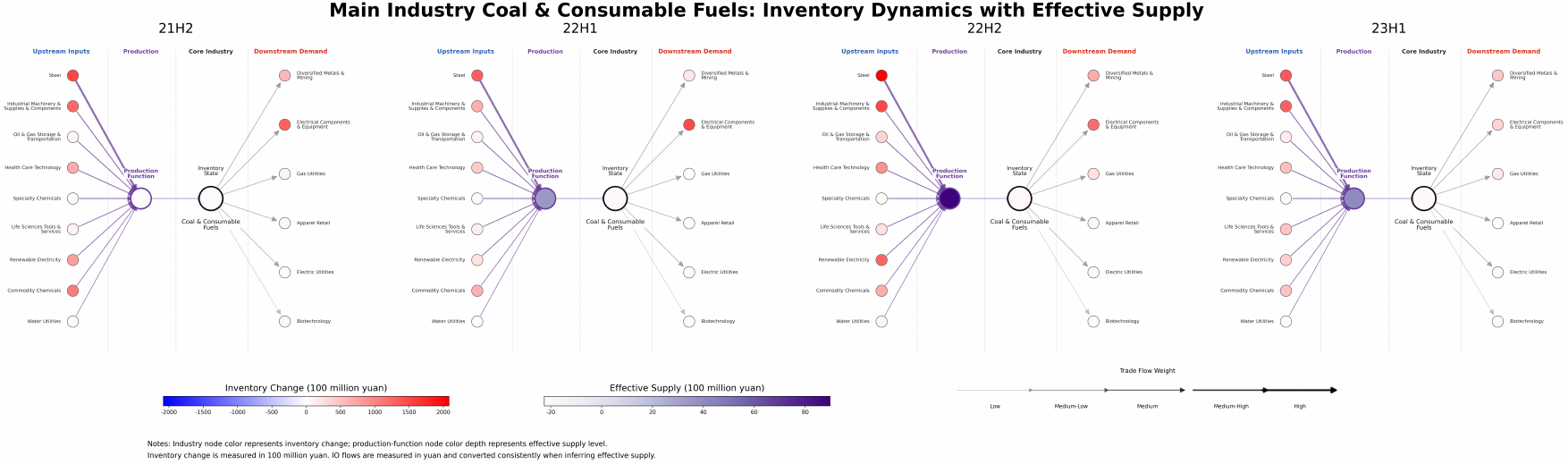}
    \caption{Visualization of Inventory Changes in Coal \& Consumable Fuels Industry Chain}
    \label{fig:Coal & Consumable Fuels}
\end{figure}

\begin{figure}
    \centering
    \includegraphics[width=1.0\linewidth]{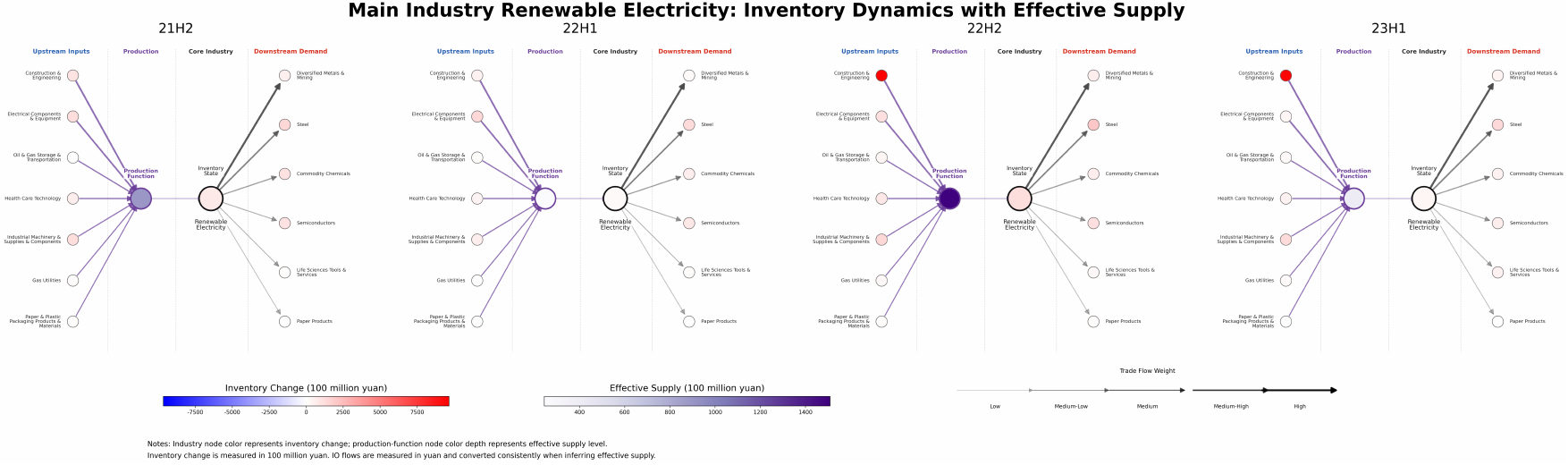}
    \caption{Visualization of Inventory Changes in Renewable Electricity Industry Chain}
    \label{fig:Renewable Electricity}
\end{figure}

Figure~\ref{fig:Coal & Consumable Fuels} and Figure~\ref{fig:Renewable Electricity} compare the traditional energy industry chain with the new energy industry chain. 
Inventory changes in them exhibit distinctly different operating mechanisms. 
The coal \& fuel industry chain displays a pattern of \emph{policy-driven} fluctuation. 
Inventory in this sector does not primarily reflect market heat; rather, it reflects the maintenance of a safety boundary. 
Constrained by long-term contract pricing and supply-assurance policies, inventory in this industry mainly functions as a \emph{shock absorber}. 
The inventory buildup during December 2021 to December 2022 can be interpreted as a form of \emph{passive defense} motivated by security considerations, 
whereas the adjustment in the fist half of 2023 represents \emph{policy-induced destocking} following the dissipation of external shocks, 
with relatively limited spillover effects along the industry chain.

By contrast, the new energy power supply chain exhibits \emph{capacity-cycle-driven} fluctuations. 
In 2022, expectations surrounding the energy transition, together with shortages of upstream components, 
pushed the entire sector into a phase of \emph{precautionary buying}. 
In 2023, however, as upstream capacity was released in a concentrated manner while downstream demand slowed, 
the industry moved rapidly from a ``rush-to-buy'' phase into a destocking phase. 
Inventory fluctuations in this sector therefore directly affect the rise and decline of related industries.

Taken together, the short-term inventory cycle in traditional energy is shaped more strongly by policy regulation and rigid constraints on resource supply, whereas the inventory cycle in new energy is driven more by production expansion and technological iteration. Although both industries are currently in an active destocking phase, traditional energy is characterized by smoother inventory adjustment following supply stabilization, whereas new energy reflects the clearing of excess capacity and inventory after an earlier phase of expansion.
\begin{figure}
    \centering
    \includegraphics[width=1.0\linewidth]{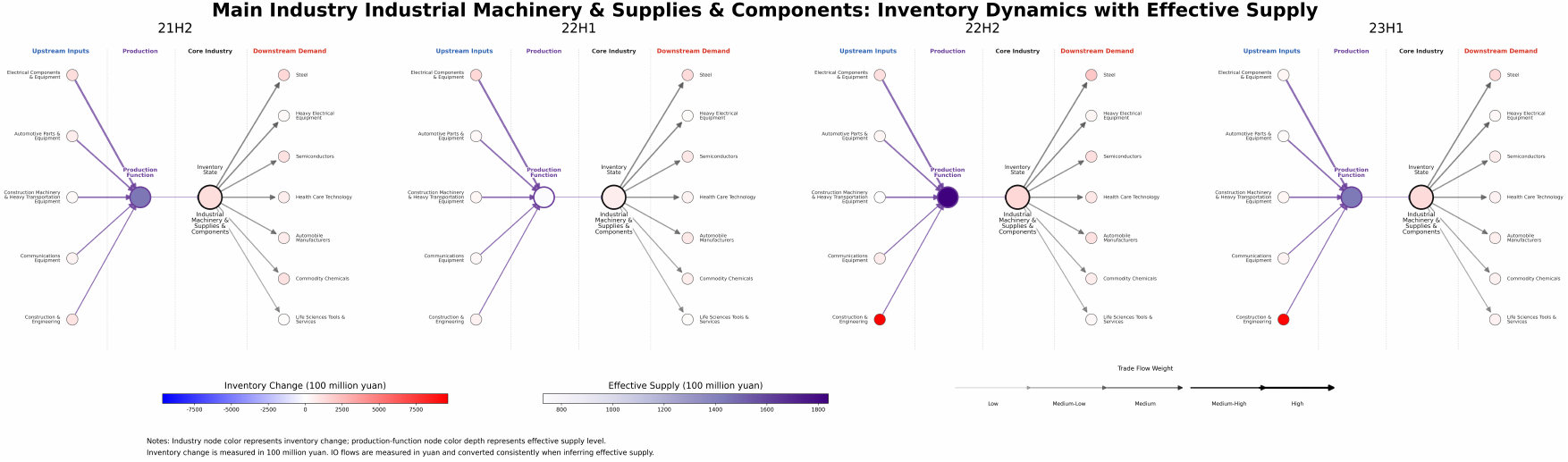}
    \caption{Visualization of Inventory Changes in Machinery \& Supplies \& Components Industry Chain}
    \label{fig:Industrial Machinery & Supplies & Components}
\end{figure}

Figure~\ref{fig:Industrial Machinery & Supplies & Components} characterizes the operating features of the industrial machinery and equipment industry. Together with the multi-agent inference results, the figure indicates that this industry exhibits typical characteristics of a midstream manufacturing sector, in which inventory changes are jointly determined by upstream input constraints, production-function-based conversion efficiency, and downstream demand for investment goods. From December 2021 to June 2022, effective supply capacity and downstream demand were broadly aligned, and inventory remained on a path of gradual expansion. In the second half of 2022, effective supply capacity strengthened noticeably, while downstream demand began to soften at the margin, causing inventory to shift toward passive accumulation. By the first half of 2023, although effective supply capacity had declined somewhat, inventory remained at a relatively high level, indicating that the industry had entered a transition from passive accumulation to active destocking. Inventory adjustment in this sector exhibits a degree of path dependence and is transmitted to both ends of the industry chain through reduced upstream procurement and a slower production pace.

Overall, the central issue currently facing the industrial machinery and equipment industry is the mismatch between earlier supply expansion and the subsequent slowdown in downstream demand. Future inventory clearance will depend on the extent to which downstream demand recovers in sectors such as steel, automobiles, semiconductors, and engineering construction.

\section{Conclusion}

Under conditions in which inventory data for SMEs are widely unavailable, we examine supply-demand mismatch and its dynamic evolution within industrial and supply chain networks. Unlike traditional analytical frameworks centered on demand forecasting or transaction flows, we treat inventory changes as the cumulative outcome of dynamic divergence between effective supply capacity and market demand, thereby representing otherwise unobservable supply-demand mismatch as an empirically tractable state variable.

Methodologically, we develop a multi-agent semi-supervised inference framework that combines prediction and validation. On the prediction side, we construct a production-function-constrained graph neural network model and use supply chain structural information to infer the effective supply capacity of firms. This allows us to predict inventory changes even when inventory data are widely unavailable. On the validation side, we further design an independent multi-agent economic consistency validation mechanism that incorporates classical econometric models to assess the economic consistency of the predictions. The results show that the predicted inventory changes not only capture contemporaneous differences between supply and demand but also remain consistent across multiple economic dimensions, indicating that the model outputs are economically meaningful.

Further analysis shows that inventory changes are not merely internal firm-level operating outcomes; rather, they reflect heterogeneous supply-demand adjustment mechanisms across different industry chains. This finding also lends support to inventory-cycle theory. For example, the coal \& fuel industry chain exhibits clear policy-driven inventory fluctuations, with inventory dynamics reflecting energy-security considerations and supply-assurance constraints. By contrast, the new energy  industry chain exhibits cyclical inventory fluctuations linked to production capacity, in which inventory changes more directly reflect the mismatch between earlier expansion and changing demand expectations. The industrial machinery and equipment industry chain displays a typical inventory-transmission pattern of midstream manufacturing, where the central tension lies in the mismatch between prior expansion in effective supply capacity and a slowdown in downstream capital expenditure; inventory clearance in this case depends on the recovery of downstream demand from sectors such as steel, automobiles, semiconductors, energy equipment, and engineering construction. Overall, the inventory prediction results across industries effectively capture heterogeneous sources of supply-demand mismatch, further indicating that our predictions are economically consistent.

Our contributions are threefold. First, we incorporate inventory changes into the analytical framework of supply-demand mismatch and represent otherwise unobservable mismatch through inventory changes driven by the gap between effective supply capacity and downstream demand. Second, we construct a production-function-constrained graph neural network model that uses supply chain network structure to infer the effective supply capacity of firms and predict inventory changes for SMEs with unavailable inventory labels. Third, we design a multi-agent economic consistency validation mechanism that evaluates whether the prediction results conform to economic regularities, thereby providing supporting evidence for the reliability of inventory predictions in unlabeled settings. Overall, our study offers a new analytical framework for understanding supply-demand mismatch in industrial and supply chain systems.


\newpage
\begingroup
\makeatletter
\def\enoteheading{}
\def\enotesize{\small}
\def\@makeenmark{\hbox{\normalfont\@theenmark.}}
\def\enoteformat{%
  \rightskip=0pt
  \leftskip=0pt
  \parindent=0pt
  \hangindent=1.5em
  \hangafter=1
  \leavevmode
  \llap{\normalfont\@theenmark.\enspace}%
}
\makeatother

\section*{Endnotes}
\theendnotes
\endgroup

\bibliographystyle{unsrtnat}  
\bibliography{references}

@article{winters1960,
  author  = {Winters, P. R.},
  title   = {Forecasting sales by exponentially weighted moving averages},
  journal = {Management Science},
  year    = {1960},
  volume  = {6},
  number  = {3},
  pages   = {324--342}
}

@article{roberts1982,
  author  = {Roberts, S. A.},
  title   = {A general class of Holt-Winters type forecasting models},
  journal = {Management Science},
  year    = {1982},
  volume  = {28},
  number  = {7},
  pages   = {808--820}
}

@article{gardner1985,
  author  = {Gardner, E. S. and McKenzie, E.},
  title   = {Forecasting trends in time series},
  journal = {Management Science},
  year    = {1985},
  volume  = {31},
  number  = {10},
  pages   = {1237--1246}
}

@article{harahap2025,
  author  = {Harahap, A. Z. M. K. and Rahim, M. K. I. A. and Malinjasari, N. and Salleh, S. M. and Maarof, R. A.},
  title   = {Enhancing inventory management through demand forecasting},
  journal = {International Journal of Research and Innovation in Social Science},
  year    = {2025},
  volume  = {9},
  number  = {1},
  pages   = {2737--2744}
}

@article{celestin2025,
  author  = {Celestin, M. and Raja, G. R. G. and Mohamed, J. A. and Kumar, D. M.},
  title   = {Mathematical models for demand forecasting in procurement: Balancing inventory and avoiding stockouts},
  journal = {International Journal of Advanced Trends in Engineering and Technology},
  year    = {2025},
  volume  = {10},
  number  = {2},
  pages   = {142--151}
}

@article{dasilva2026,
  author  = {da Silva, L. F. and Christo, E. S. and Costa, K. A. and Tavares, H. P.},
  title   = {Combining demand classification and forecasting models in spare parts inventory management for the energy sector},
  journal = {International Journal of Advanced Manufacturing Technology},
  year    = {2026}
}

@article{olaniyi2026,
  author  = {Olaniyi, O. A. and Pugal, P. S. and Etim, M.},
  title   = {Optimising inventory management strategies for cost reduction in supply chains: A systematic review},
  journal = {The Sankalpa: International Journal of Management Decisions},
  year    = {2026},
  volume  = {12},
  number  = {1},
  pages   = {97--103}
}

@article{sagaert2025,
  author  = {Sagaert, Y. R. and Kourentzes, N.},
  title   = {Inventory management with leading indicator augmented hierarchical forecasts},
  journal = {Omega},
  year    = {2025},
  volume  = {136},
  pages   = {103335}
}

@article{zizka2026,
  author  = {Zizka, M. and Kustka, M.},
  title   = {Methodology for inventory management with sporadic demand in small and medium-sized enterprises},
  journal = {Acta Logistica},
  year    = {2026},
  volume  = {13},
  number  = {1},
  pages   = {174--185}
}

@article{acemoglu2012,
  author  = {Acemoglu, D. and Carvalho, V. M. and Ozdaglar, A. and Tahbaz-Salehi, A.},
  title   = {The network origins of aggregate fluctuations},
  journal = {Econometrica},
  year    = {2012},
  volume  = {80},
  number  = {5},
  pages   = {1977--2016}
}

@article{carvalho2021,
  author  = {Carvalho, V. M. and Nirei, M. and Saito, Y. U. and Tahbaz-Salehi, A.},
  title   = {Supply chain disruptions: Evidence from the Great East Japan Earthquake},
  journal = {The Quarterly Journal of Economics},
  year    = {2021},
  volume  = {136},
  number  = {2},
  pages   = {1255--1321}
}

@article{wu2023,
  author  = {Wu, D. and Wang, Q. and Olson, D. L.},
  title   = {Industry classification based on supply chain network information using graph neural networks},
  journal = {Applied Soft Computing},
  year    = {2023},
  volume  = {132}
}

@inproceedings{huang2023,
  author    = {Huang, S. and Poursafaei, F. and Danovitch, J. and Fey, M. and Hu, W. and Rossi, E. and Leskovec, J. and Bronstein, M. and Rabusseau, G. and Rabbany, R.},
  title     = {Temporal graph benchmark for machine learning on temporal graphs},
  booktitle = {Proceedings of the 37th Annual Conference on Neural Information Processing Systems (NeurIPS'23)},
  year      = {2023}
}

@article{tu2024,
  author  = {Tu, Y. and Li, W. and Song, X. and Gong, K. and Liu, L. and Qin, Y. and Liu, Y. and Liu, M.},
  title   = {Using graph neural network to conduct supplier recommendation based on large-scale supply chain},
  journal = {International Journal of Production Research},
  year    = {2024},
  pages   = {1--14}
}

@inproceedings{tan2022,
  author    = {Tan, Z. and Liu, B. and Yin, G.},
  title     = {Asymmetric self-supervised graph neural networks},
  booktitle = {2022 IEEE International Conference on Big Data (Big Data)},
  year      = {2022},
  pages     = {1369--1376},
  publisher = {IEEE}
}

@article{kitchin1923,
  author  = {Kitchin, J.},
  title   = {Cycles and trends in economic factors},
  journal = {Review of Economics and Statistics},
  year    = {1923},
  volume  = {5},
  number  = {1},
  pages   = {10--16}
}

@article{metzler1941,
  author  = {Metzler, L. A.},
  title   = {The nature and stability of inventory cycles},
  journal = {Review of Economics and Statistics},
  year    = {1941},
  volume  = {23},
  number  = {3},
  pages   = {113--129}
}

@article{blinder1991,
  author  = {Blinder, A. S. and Maccini, L. J.},
  title   = {Taking stock: A critical assessment of recent research on inventories},
  journal = {Journal of Economic Perspectives},
  year    = {1991},
  volume  = {5},
  number  = {1},
  pages   = {73--96}
}

@article{bils2000,
  author  = {Bils, M. and Kahn, J. A.},
  title   = {What inventory behavior tells us about business cycles},
  journal = {American Economic Review},
  year    = {2000},
  volume  = {90},
  number  = {3},
  pages   = {458--481}
}

@article{lee1997,
  author  = {Lee, Hau L. and Padmanabhan, V. and Whang, Seungjin},
  title   = {Information Distortion in a Supply Chain: The Bullwhip Effect},
  journal = {Management Science},
  year    = {1997},
  volume  = {43},
  number  = {4},
  pages   = {546--558},
  doi     = {10.1287/mnsc.43.4.546}
}

@article{osadchiy2021,
  author  = {Osadchiy, Nikolay and Schmidt, William and Wu, Jing},
  title   = {The Bullwhip Effect in Supply Networks},
  journal = {Management Science},
  year    = {2021},
  volume  = {67},
  number  = {10},
  pages   = {6153--6173},
  doi     = {10.1287/mnsc.2020.3824}
}

@article{qu2021,
  author  = {Qu, Zhan and Raff, Horst},
  title   = {Vertical Contracts in a Supply Chain and the Bullwhip Effect},
  journal = {Management Science},
  year    = {2021},
  volume  = {67},
  number  = {6},
  pages   = {3744--3756},
  doi     = {10.1287/mnsc.2020.3630}
}

@article{chu2017,
  author  = {Chu, Leon Yang and Shamir, Noam and Shin, Hyoduk},
  title   = {Strategic Communication for Capacity Alignment with Pricing in a Supply Chain},
  journal = {Management Science},
  year    = {2017},
  volume  = {63},
  number  = {12},
  pages   = {4366--4388},
  doi     = {10.1287/mnsc.2016.2527}
}

@article{candogan2024,
  author  = {Candogan, Ozan and Gurkan, Huseyin},
  title   = {The Value of Information Design in Supply Chain Management},
  journal = {Management Science},
  year    = {2025},
  volume  = {71},
  number  = {8},
  pages   = {6545--6558},
  doi     = {10.1287/mnsc.2023.03004}
}

@article{qi2022,
  author  = {Qi, Meng and Shi, Yuanyuan and Qi, Yongzhi and Ma, Chenxin and Yuan, Rong and Wu, Di and Shen, Zuo-Jun Max},
  title   = {A Practical End-to-End Inventory Management Model with Deep Learning},
  journal = {Management Science},
  year    = {2022},
  volume  = {69},
  number  = {2},
  pages   = {759--773},
  doi     = {10.1287/mnsc.2022.4564}
}

@article{li2023,
  author  = {Li, Xiaocheng and Zheng, Zeyu},
  title   = {Dynamic Pricing with External Information and Inventory Constraint},
  journal = {Management Science},
  year    = {2023},
  volume  = {70},
  number  = {9},
  pages   = {5985--6001},
  doi     = {10.1287/mnsc.2023.4963}
}

@article{chen2024,
  author  = {Chen, Boxiao and Jiang, Jiashuo and Zhang, Jiawei and Zhou, Zhengyuan},
  title   = {Learning to Order for Inventory Systems with Lost Sales and Uncertain Supplies},
  journal = {Management Science},
  year    = {2024},
  volume  = {70},
  number  = {12},
  pages   = {8631--8646},
  doi     = {10.1287/mnsc.2022.02476}
}

@article{zhang2025,
  author  = {Zhang, Kairen and Gao, Xiangyu and Wang, Zhanyue and Zhou, Sean X.},
  title   = {Sampling-Based Approximation for Series Inventory Systems},
  journal = {Management Science},
  year    = {2025},
  volume  = {71},
  number  = {10},
  pages   = {8200--8217},
  doi     = {10.1287/mnsc.2022.01876}
}

@article{wu_lai2022,
  author  = {Wu, Qi and Lai, Guoming},
  title   = {The Effects of Stock-Based Incentives on Inventory Management},
  journal = {Management Science},
  year    = {2022},
  volume  = {68},
  number  = {7},
  pages   = {5068--5086},
  doi     = {10.1287/mnsc.2021.4087}
}

@article{hsu_wu2024,
  author  = {Hsu, Vernon and Wu, Jing},
  title   = {Inventory as a Financial Instrument: Evidence from China's Metal Industries},
  journal = {Management Science},
  year    = {2024},
  volume  = {70},
  number  = {6},
  pages   = {3645--3663},
  doi     = {10.1287/mnsc.2023.4873}
}

@article{iancu2017,
  author  = {Iancu, Dan A. and Trichakis, Nikolaos and Tsoukalas, Gerry},
  title   = {Is Operating Flexibility Harmful Under Debt?},
  journal = {Management Science},
  year    = {2017},
  volume  = {63},
  number  = {6},
  pages   = {1730--1761},
  doi     = {10.1287/mnsc.2015.2415}
}

@article{barrot2016,
  author  = {Barrot, Jean-No{\"e}l and Sauvagnat, Julien},
  title   = {Input Specificity and the Propagation of Idiosyncratic Shocks in Production Networks},
  journal = {Quarterly Journal of Economics},
  year    = {2016},
  volume  = {131},
  number  = {3},
  pages   = {1543--1592},
  doi     = {10.1093/qje/qjw018}
}

@article{antras2012,
  author  = {Antr{\`a}s, Pol and Chor, Davin and Fally, Thibault and Hillberry, Russell},
  title   = {Measuring the Upstreamness of Production and Trade Flows},
  journal = {American Economic Review},
  year    = {2012},
  volume  = {102},
  number  = {3},
  pages   = {412--416},
  doi     = {10.1257/aer.102.3.412}
}

@article{chehrazi2025,
  author  = {Chehrazi, N.},
  title   = {Inventory Systems with Record Inaccuracy: Transaction Errors vs. Unobservable Loss},
  journal = {Manufacturing \& Service Operations Management},
  volume  = {27},
  number  = {4},
  pages   = {1183--1204},
  year    = {2025}
}

@article{iglehart1972,
  author  = {Iglehart, D. L. and Morey, R. C.},
  title   = {Inventory Systems with Imperfect Asset Information},
  journal = {Management Science},
  volume  = {18},
  number  = {8},
  pages   = {B388--B394},
  year    = {1972}
}

@article{dehoratius2008,
  author  = {DeHoratius, N. and Raman, A.},
  title   = {Inventory Record Inaccuracy: An Empirical Analysis},
  journal = {Management Science},
  volume  = {54},
  number  = {4},
  pages   = {627--641},
  year    = {2008}
}

@article{SerpaKrishnan2018,
  author  = {Serpa, Juan Carlos and Krishnan, Harish},
  title   = {The Impact of Supply Chains on Firm-Level Productivity},
  journal = {Management Science},
  volume  = {64},
  number  = {2},
  pages   = {511--532},
  year    = {2018}
}

@article{WangLiWuAnupindi2021,
  author  = {Wang, Yimin and Li, Jing and Wu, Di and Anupindi, Ravi},
  title   = {When Ignorance Is Not Bliss: An Empirical Analysis of Subtier Supply Network Structure on Firm Risk},
  journal = {Management Science},
  volume  = {67},
  number  = {4},
  pages   = {2029--2048},
  year    = {2021}
}

@article{BimpikisCandoganEhsani2019,
  author  = {Bimpikis, Kostas and Candogan, Ozan and Ehsani, Shayan},
  title   = {Supply Disruptions and Optimal Network Structures},
  journal = {Management Science},
  volume  = {65},
  number  = {12},
  pages   = {5504--5517},
  year    = {2019}
}

@article{FariasLiPeng2024,
  author  = {Farias, Vivek F. and Li, Andrew A. and Peng, Tianyu},
  title   = {Fixing Inventory Inaccuracies at Scale},
  journal = {Manufacturing \& Service Operations Management},
  volume  = {26},
  number  = {2},
  pages   = {1102--1118},
  year    = {2024}
}

@article{agrawal2023,
  author  = {Agrawal, Deepak and Osadchiy, Nikolay},
  title   = {Inventory Productivity and Stock Returns in Manufacturing Networks},
  journal = {Manufacturing \& Service Operations Management},
  year    = {2023},
  volume  = {26},
  number  = {2},
  pages   = {573--593},
  doi     = {10.2139/ssrn.4166155}
}

@article{baqaee2020,
  author  = {Baqaee, David R. and Farhi, Emmanuel},
  title   = {Supply and Demand in Disaggregated Keynesian Economies with an Application to the COVID-19 Crisis},
  journal = {NBER Working Paper},
  year    = {2020},
  number  = {27152},
  doi     = {10.3386/w27152}
}

@article{baqaee2024,
  author  = {Baqaee, David R. and Farhi, Emmanuel},
  title   = {Networks, Barriers, and Trade},
  journal = {Econometrica},
  year    = {2024},
  volume  = {92},
  number  = {2},
  pages   = {505--541},
  doi     = {10.3982/ECTA19430}
}

@article{chiang2023,
  title     = {Examining demand and supply-chain antecedents of inventory dynamics: Evidence from automotive industry},
  author    = {Chiang, Chung-Yean and Qian, Zhuang and Chuang, Chia-Hung and Tang, Xiao and Chou, Chia-Ching},
  journal   = {International Journal of Production Economics},
  volume    = {259},
  pages     = {108838},
  year      = {2023},
  publisher = {Elsevier},
  doi       = {10.1016/j.ijpe.2023.108838}
}

@article{wang2021,
  author  = {Wang, Xinyu and Lin, Yong and Shi, Yina},
  title   = {Linking Industrial Agglomeration and Manufacturers Inventory Performance: The Moderating Role of Firm Size and Enterprise Status in the Supply Chain},
  journal = {Journal of Manufacturing Technology Management},
  year    = {2021},
  volume  = {32},
  number  = {2},
  pages   = {448--484},
  doi     = {10.1108/JMTM-03-2020-0085}
}

@article{larson2015,
  title   = {An Empirical Investigation of Dynamic Ordering Policies},
  author  = {Larson, Chad R. and Turcic, Danko and Zhang, Fuqiang},
  journal = {Management Science},
  volume  = {61},
  number  = {9},
  pages   = {2118--2138},
  year    = {2015}
}

@article{KOSCHAT2008,
  title   = {Store inventory can affect demand: Empirical evidence from magazine retailing},
  journal = {Journal of Retailing},
  volume  = {84},
  number  = {2},
  pages   = {165-179},
  year    = {2008},
  issn    = {0022-4359},
  doi     = {https://doi.org/10.1016/j.jretai.2008.04.003},
  url     = {https://www.sciencedirect.com/science/article/pii/S0022435908000146},
  author  = {Martin A. Koschat}
}

\end{document}